\documentclass[12pt]{article}
\usepackage{amsmath,amssymb,graphicx} 
\usepackage{epsf}
\usepackage{epsfig}
\usepackage{pstricks}
\usepackage{cite}

\newcommand{\be}{\begin{equation}}
\newcommand{\ee}{\end{equation}}

\newcommand{\beq}{\begin{eqnarray}}
\newcommand{\eeq}{\end{eqnarray}}

\newcommand{\centeron}[2]{{\setbox0=\hbox{#1}\setbox1=\hbox{#2}\ifdim
\wd1>\wd0\kern.5\wd1\kern-.5\wd0\fi
\copy0

\kern-.5\wd0\kern-.5\wd1\copy1\ifdim\wd0>\wd1
                                       \kern.5\wd0\kern-.5\wd1\fi}}
\newcommand{\ltap}{\>\centeron{\raise.35ex\hbox{$<$}}
                               {\lower.65ex\hbox{$\sim$}}\>}
\newcommand{\gtap}{\>\centeron{\raise.35ex\hbox{$>$}}
                               {\lower.65ex\hbox{$\sim$}}\>}

\newcommand\ZZ{\hbox{\zfont Z\kern-.4emZ}}
\font\zfont = cmss10 

\begin{document}
\begin{titlepage}
\begin{flushright}
\end{flushright}

\vskip.5cm
\begin{center}
{\Large \bf 
Electroweak and finite quark-mass effects on the Higgs boson transverse momentum distribution
}

\vskip.1cm
\end{center}
\vskip0.2cm

\begin{center}
{\bf
Wai-Yee Keung}
\end{center}

\begin{center}
{\it Physics Department, University of Illinois, Chicago, IL 60607} \\
\vspace*{0.1cm}
\vspace*{0.3cm}
{\tt keung@uic.edu}
\end{center}

\begin{center}
{\bf
Frank J. Petriello}
\end{center}

\begin{center}
{\it Physics Department, University of Wisconsin, WI 53706} \\

\vspace*{0.1cm}
\vspace*{0.3cm}
{\tt frankjp@physics.wisc.edu}
\end{center}

\vglue 0.3truecm

\begin{abstract}
\vskip 3pt
\noindent

We perform a detailed study of the various one-loop contributions leading to production of the Standard Model Higgs boson in association with a hard jet.  This production mode contributes to the current Tevatron exclusion limit of the Standard Model Higgs with $160\,\,{\rm GeV} \leq M_H \leq 170\,\,{\rm GeV}$, and will also be important for discovery and interpretation of new scalar bosons at the 
Large Hadron Collider (LHC).  We include top- and bottom-quark initiated contributions, maintaining the exact dependence on the quark masses, and also study previously neglected $W$- and $Z$-boson mediated 
effects which shift the $qg$ and $q\bar{q}$ production modes.  We consider the deviations from commonly used approximations for the Higgs boson transverse momentum spectrum caused 
by the finite top-quark mass, bottom quark contributions, and electroweak gauge boson terms.  All three effects act to decrease the Higgs boson transverse momentum distribution for observable momenta, with shifts reaching $-8\%$ at the Tevatron and $-30\%$ at the LHC.  The shifts have a significant dependence on the Higgs $p_T$, and are especially important if large momenta are selected by experimental cuts. 

\end{abstract}

\end{titlepage}

\newpage


\section{Introduction}
\label{sec:intro}

The Higgs boson is the last undiscovered particle of the Standard Model (SM), and its discovery is a major goal of both the Tevatron and the Large Hadron Collider (LHC) physics programs.  Direct searches at LEP restrict the SM Higgs boson to have a mass $M_H>114.4$ GeV~\cite{Barate:2003sz}.  Recently, the Tevatron collaborations have ruled out the mass range $160\,{\rm GeV} \leq M_H \leq 170 \,{\rm GeV}$ at 
95\% C.L.~\cite{TEVhiggs}.  If a scalar particle is discovered at either collider, the measurement of its properties will be crucial to determine whether the particle found is the Standard Model Higgs boson, 
or whether it hints at physics beyond the SM.

The theoretical community has devoted significant effort to understanding precisely the production cross section and decay widths of the SM Higgs particle in order to facilitate such studies.  The dominant production mode at both the Tevatron and the LHC is the partonic mechanism $gg\to H$ proceeding through a top-quark loop~\cite{Ellis:1975ap,Georgi:1977gs,Wilczek:1977zn,Shifman:1979eb,Rizzo:1980gz}.  The exact next-to-leading order (NLO) QCD corrections to this process are known, including the corrections to the smaller bottom-quark induced contribution~\cite{Djouadi:1991tka,Spira:1995rr}.  In the effective theory with the 
top quark integrated out by taking $m_t \to \infty$ to produce an $HGG$ operator, both the NLO corrections~\cite{Dawson:1990zj} and the NNLO corrections are known~\cite{Harlander:2002wh,Anastasiou:2002yz,Ravindran:2003um}.  When normalized 
to the full $m_t$-dependent leading-order result, this effective theory reproduces the exact NLO result to better than 1\% for $M_H < 2m_t$ and to 10\% or better for Higgs boson masses up to 1 TeV~\cite{ztalks}.  The NNLO differential distributions in the effective theory were first obtained in Refs.~\cite{Anastasiou:2004xq,Anastasiou:2005qj,Catani:2007vq,Catani:2008me}, and detailed studies of the effects of experimental cuts on Higgs boson cross sections have been performed~\cite{Davatz:2006ut,Anastasiou:2007mz,Anastasiou:2008ik,Grazzini:2008tf}.  Resummation of logarithmically-enhanced threshold corrections to the cross section has been studied~\cite{Catani:2003zt,Ahrens:2008qu,Ahrens:2008nc}.  It was recently demonstrated that corrections arising from the Higgs coupling to $W$- and $Z$-bosons could affect the $gg \to H$ cross section at the 6-7\% level~\cite{Aglietti:2004nj,Actis:2008ug,Anastasiou:2008tj}.  The inclusion of such corrections is important for precision predictions of Higgs boson properties, and also impacts the exclusion 
limits set by the Tevatron collaborations.  The theoretical uncertainty from missing higher-order terms is estimated to be roughly $\pm 10\%$ after these QCD and electroweak effects are accounted for.  

Production of a Higgs boson in association with one or more jets is an important process for experimental searches.  For example, the study of $pp \to H+{\rm jets}$ with the subsequent decay 
$H \to \gamma\gamma$~\cite{Abdullin:1998er} or $H \to WW$~\cite{Mellado:2007fb} has significantly different kinematics than its background, and offers a good potential for Higgs discovery at the LHC.  At the Tevatron, the multi-jet bins contribute roughly 50\% of the exclusion limit~\cite{TEVmultijet}.  Since the Higgs boson obtains a transverse momentum from recoil against hard jets, the study 
of Higgs+jet production is related to that of the Higgs transverse momentum spectrum.  The calculation of Higgs boson production in association with a jet proceeding through top-quark loops, maintaining the exact dependence on the mass of the internal quark, was first presented in Ref.~\cite{Ellis:1987xu} and further studied in Ref.~\cite{Baur:1989cm}.  The NLO QCD corrections to the Higgs boson transverse momentum spectrum have been computed in the effective theory with $m_t \to \infty$~\cite{deFlorian:1999zd,Ravindran:2002dc,Glosser:2002gm}, and the effect of bottom-quark induced contributions has been quantified~\cite{Field:2003yy}.  The resummation of logarithmically enhanced corrections 
in the low $p_T$ limit have been considered~\cite{Kauffman:1991jt,Yuan:1991we,Berger:2002ut,Bozzi:2003jy,Bozzi:2005wk}.  In addition to aiding the discovery of the Higgs boson, study of its transverse momentum 
distribution has been discussed as a probe of physics beyond the SM~\cite{Langenegger:2006wu,Brein:2003df,Brein:2007da,Arnesen:2008fb}.

In light of the current experimental sensitivity at the Tevatron to Higgs production in association with a jet, and also in preparation for future studies at the LHC, we perform a detailed study of the various contributions to the one-loop 
Higgs transverse momentum spectrum in the SM.  Some of these effects have previously been studied separately; we combine them into a single calculation and consider their collective impact.  We study the effect of the top-quark mass and bottom quarks on the $gg\to Hg$, $qg\to Hq$, and $q\bar{q} \to Hg$ partonic 
processes.  Since electroweak effects were shown to have a sizable effect on the inclusive cross section, we also consider $W$ and $Z$ boson mediated electroweak contributions to the Higgs $p_T$ spectrum.  These affect both the $q\bar{q}$ and $qg$ partonic processes, and were first considered in Ref.~\cite{Mrenna:1995cf}.  However, the interference of these such corrections with the top-quark initiated diagrams, which we find to be the dominant effect, was not studied.  For observable values of the Higgs transverse momentum at both the Tevatron and the LHC, the effects of the finite top-quark mass, bottom quark terms, and electroweak terms all act destructively to reduce the high $p_T$ cross section obtained from the $m_t \to \infty$ effective theory.  For relevant momenta at the Tevatron, the reduction reaches $-8\%$.  At the LHC, the reduction can reach $-30\%$ for Higgs transverse momenta with 
appreciable rates.  We combine the exact $t$, $b$, $W$, and $Z$ mediated contributions in a {\tt FORTRAN} code which allows arbitrary cuts on the Higgs boson kinematics to be imposed.

Our paper is organized as follows.  In Section~\ref{sec:calc} we present details of our analytic computation of the various terms which give rise to the Higgs $p_T$ distribution.  We present numerical 
results for the Tevatron and for the LHC with $\sqrt{s}=10$ TeV in Section~\ref{sec:numerics}.  We conclude in Section~\ref{sec:conc}.  

\section{Calculational details}
\label{sec:calc}

\begin{figure}[htbp]
   \centering
   \includegraphics[width=0.70\textwidth,angle=0]{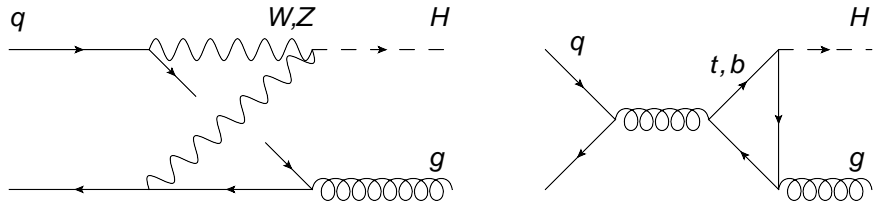}
   \caption{Example one-loop diagrams contributing the the partonic process $q\bar{q} \to Hg$ through internal $W,Z$ bosons (left figure) or fermions (right figure).}
   \label{qqbdiags}
\end{figure}

We describe our calculation of the Higgs boson transverse momentum spectrum at one-loop accuracy, including terms induced by both heavy quarks and electroweak gauge bosons.  We 
begin with the partonic channels affected by both quarks and electroweak corrections, the $q\bar{q}$ and $qg$ initial states.  Example diagrams contributing to the partonic processes $q\bar{q} \to Hg$ and $qg \to Hq$ are shown in Fig.~\ref{qqbdiags}.  As the two processes are related by a crossing symmetry, we demonstrate the calculational details for the process $q(p_1)+\bar{q}(p_2) \to H(p_H)+g(p_3)$.  The amplitude for this process can be expanded in terms of two spin structures and their associated form factors,
\begin{equation}
{\cal M} = \frac{1}{16\pi^2}\sum_{x=W,Z,b,t} {\cal F}^x(\hat{t},\hat{u},\hat{s},M_H,m_x) \Gamma^x_1 + {\cal F}^x(\hat{u},\hat{t},\hat{s},M_H,m_x) \Gamma^x_2,
\label{amp}
\end{equation}
where we have defined the Mandelstam invariants $\hat{t}=(p_1-p_3)^2$, $\hat{u}=(p_2-p_3)^2$, $\hat{s}=(p_1+p_2)^2$.  The sum over $x$ includes all internal states with relevant couplings to the Higgs boson.  We include $W$ and $Z$ bosons, top quarks, and bottom quarks.  The spin structures can be written as 
\begin{eqnarray}
\Gamma^x_1 &=& T^a_{ij} \left[p_1 \cdot p_3 \,\bar{v}_2^i \slash\!\!\!{\epsilon}_3^a (v_x+a_x \gamma_5)u_1^j - p_1 \cdot \epsilon_3^a \,\bar{v}_2^i \slash\!\!\!{p}_3 (v_x+a_x \gamma_5)u_1^j\right], \nonumber \\
\Gamma^x_2 &=& T^a_{ij} \left[p_2 \cdot p_3 \,\bar{v}_2^i \slash\!\!\!{\epsilon}_3^a (v_x+a_x \gamma_5)u_1^j - p_2 \cdot \epsilon_3^a \,\bar{v}_2^i \slash\!\!\!{p}_3 (v_x+a_x \gamma_5)u_1^j\right].
\label{spinstrucs}
\end{eqnarray}
We note that this process conserves chirality, and is not suppressed by the initial-state light-quark masses.  The relevant couplings are simple to derive in terms of standard Feynman rules, and are found to be as follows:
\begin{eqnarray}
v_W &=& g_s g M_W \left[(w_v^{q})^2+(w_a^{q})^2\right],\,\,\, a_W = 2g_s g M_W w_v^q w_a^q,\,\,\,v_Z = g_s \frac{g}{c_W} M_Z \left[(z_v^{q})^2+(z_a^{q})^2\right],
\nonumber \\
a_Z &=& 2 g_s\frac{g}{c_W}M_Z z_v^q z_a^q,\,\,\, v_{t,b} = -g_s^3 g \frac{m_{t,b}}{2M_W},\,\,\, a_{t,b}=0,\,\,\,w_v^u=w_v^d=\frac{g}{2\sqrt{2}},
\nonumber \\
w_a^u&=&w_a^d=-\frac{g}{2\sqrt{2}},\,\,\, z_v^u= \frac{g}{c_W}\left(\frac{1}{4}-\frac{2}{3}s_W^2\right),\,\,\, z_a^u=-\frac{g}{4c_W}, \nonumber \\
z_v^d&=& \frac{g}{c_W}\left(-\frac{1}{4}+\frac{1}{3}s_W^2\right),\,\,\, z_a^d=\frac{g}{4c_W}.
\end{eqnarray}
Here, $g_s$ and $g$ are respectively the strong and weak coupling constants, and $c_W$, $s_W$ are the cosine and sine of the weak mixing angle.  We have parametrized the couplings of the quarks to the $W,Z$ bosons as $i \left(w_v+w_a\gamma_5\right)$ and $i \left(z_v+z_a\gamma_5\right)$, respectively.  For 
$w^x_{v,a}$ and $z^x_{v,a}$ the superscripts $u$ and $d$ denote respectively arbitrary up- and down-type quarks.

One interesting fact to note about the amplitude defined in Eqs.~(\ref{amp}) and~(\ref{spinstrucs}) is that it vanishes when $p_3$ becomes proportional to 
either $p_1$ or $p_2$.  As the electroweak corrections have no contribution from $t$-channel gluon exchange, this suggests that they will have a harder $p_T$ spectrum than the typical contributions, which is indeed what we 
find later.  To derive expressions for the form factors ${\cal F}^x$, we use standard techniques to reduce the contributing diagrams to a set of master integrals.  We implement this 
reduction using the packages FORM~\cite{Vermaseren:2000nd} and QGRAF~\cite{Nogueira:1991ex}.  We give analytic expressions for the form factors in the Appendix.  We note that the expressions for the top and bottom quark were first derived in Ref.~\cite{Ellis:1987xu}, but we reproduce them here for completeness.

Given the expression for the amplitude, it is straightforward to derive the differential cross section.  In terms of the partonic center-of-momentum frame scattering angle $\theta$
of the final-state Higgs boson, the differential cross section is
\begin{equation}
\frac{d \sigma}{d\, {\rm cos}\theta} = \frac{1}{36} \frac{1}{2\hat{s}}\frac{(1-M_H^2/\hat{s})}{16\pi} \sum_{spin,color} |{\cal M}|^2,
\end{equation}
with $\hat{s}+\hat{t}+\hat{u}=M_H^2$.  The differential result for the $qg$ channel can be obtained by crossing $\hat{t} \leftrightarrow \hat{s}$ and by changing the initial-state color averaging.

\begin{figure}[htbp]
   \centering
   \includegraphics[width=0.70\textwidth,angle=0]{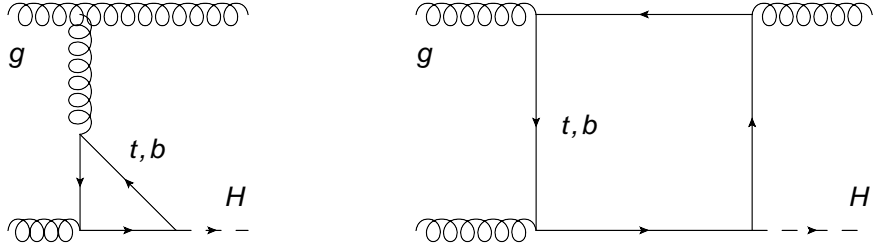}
   \caption{Example one-loop diagrams contributing the the partonic process $gg \to Hg$ through internal top and bottom quarks.}
   \label{ggdiags}
\end{figure}

We also study the numerical effect of bottom quark and $m_t \neq \infty$ corrections to the dominant partonic process leading to production of a Higgs boson plus one jet, 
$g(p_1)+g(p_2) \to H+g(p_3)$.  We have reproduced the cross section for this partonic process, which was first derived in Ref.~\cite{Ellis:1987xu}.  Example diagrams are shown in Fig.~\ref{ggdiags}.  The amplitude for this process takes the form
\begin{equation}
{\cal M}_{gg} = - \frac{g_s^3 g}{16\pi^2} \sum_i \sum_{x=b,t} \frac{m_{x}}{2M_W} {\cal G}_i(\hat{t},\hat{u},\hat{s},M_H,m_x) \Gamma_i,
\label{ampgg}
\end{equation}
with the same definition of Mandelstam invariants as utilized above for the other channels.  We choose the following basis of spin structures to describe this process:
\begin{eqnarray}
\Gamma_1 &=& \epsilon_1^a \cdot \epsilon_2^b\, p_2 \cdot \epsilon_3^c f^{abc},\,\,\, \Gamma_2 = \epsilon_2^b \cdot \epsilon_3^c \,p_3 \cdot \epsilon_1^a f^{abc}, \nonumber \\
\Gamma_3 &=& \epsilon_1^a \cdot \epsilon_3^c\, p_3 \cdot \epsilon_2^b f^{abc},\,\,\, \Gamma_4 = p_3 \cdot \epsilon_1^a \,p_3 \cdot \epsilon_2^b \,p_2 \cdot \epsilon_3^c f^{abc}.
\end{eqnarray}
When computing $|{\cal M}|^2$, we use the following sums over the various polarizations:
\begin{equation}
\sum_{pols} \epsilon_x^{\mu} (\epsilon_x^{\nu})^{*} = -g^{\mu \nu} +\frac{p_x^{\mu} n_x^{\nu} +p_x^{\nu} n_x^{\mu}}{p_x \cdot n_x},
\end{equation} 
with $n_1=p_2$, $n_2=p_1$, and $n_3=p_1$.  The expressions for the form factors ${\cal G}_i$ are again straightforward to derive, and are given in the Appendix.  The differential cross section for the $gg$ partonic channel takes the form 
\begin{equation}
\frac{d \sigma}{d\, {\rm cos}\theta} = \frac{1}{256} \frac{1}{2\hat{s}}\frac{(1-M_H^2/\hat{s})}{16\pi} \sum_{spin,color} |{\cal M}_{gg}|^2.
\end{equation}

All amplitudes considered in this paper are finite and do not require renormalization.  We have checked both analytically and numerically that our exact heavy-quark mediated contributions reproduce the effective theory results upon expansion in the fermion mass.  We have written a {\tt FORTRAN} code implementing the heavy-quark and electroweak contributions to Higgs+jet production valid for arbitrary Higgs boson mass, and which allows phase-space constraints to be imposed.  We now use this code to study the 
numerical impact of the electroweak corrections, the bottom quark, and finite-$m_t$ effects, on the Higgs transverse momentum spectrum at the Tevatron and the LHC.

\section{Numerics}
\label{sec:numerics}

We present numerical results to quantify the effect of various contributions to the Higgs transverse momentum spectrum.  We utilize the parameter values that follow:
\begin{eqnarray}
M_W &=& 80.399\,{\rm GeV},\,\,\, M_Z = 91.188\,{\rm GeV}, \nonumber \\
m_t &=& 172.4\,{\rm GeV},\,\,\, m_b = 4.8\,{\rm GeV}, \nonumber \\
G_F &=& 1.16639 \times 10^{-5}\, {\rm GeV}^{-2}.
\end{eqnarray}
To define the remaining electroweak parameters we choose $M_W$, $M_Z$, and $G_F$ as inputs and employ tree-level relations between the various quantities.  We use MSTW 2008 parton distribution functions~\cite{Martin:2009iq} extracted to NLO accuracy, and the consistent choice of $\alpha_s$ with NLO running.  For 
the studies we perform here we set $\mu_R=\mu_F=M_H$.  We note that the $b$-quark contributions are small enough that our use of the pole mass versus a running mass has very little effect on the numerical results.  We set $\sqrt{s}=1.96\,{\rm TeV}$ for the Tevatron center-of-mass energy, and $\sqrt{s}=10\,{\rm TeV}$ for the 
LHC.  To illustrate the sizes of the considered effects we set $M_H=120$ GeV.

We compute all cross sections at the one-loop level, which are the leading contributions to the Higgs $p_T$ spectrum.  The next-to-leading order corrections to the 
top-quark contributions are known in the limit $m_t \to \infty$, but corrections to the $W$, $Z$, and bottom-quark terms we also consider are not known.  Our primary 
results are the fractional deviations induced by electroweak effects, $b$-quarks and the finite top-quark mass.  If the corrections to the various terms are similar to that of the top quark, then the fractional deviations found here are unchanged.  Since most of the effects are destructive interferences which reduce the rate, using the deviations found here leads to a conservative estimate if the corrections to the $W$, $Z$, and $b$ terms are smaller.  For the top quark, an alternative approach is to attempt to include the effective theory $K$-factor to estimate the NLO corrections to the $p_T$ spectrum~\cite{Smith:2005yqa}. 

\subsection{Tevatron}

We begin by discussing the numerical effects of the various terms on the transverse momentum spectrum at the Tevatron.  The dominant partonic process leading to production of a 
Higgs boson in association with a jet is $gg \to Hg$, which occurs through the Higgs coupling to top and bottom quarks.  The contributions from the $qg$ channel contributes roughly half as much as $gg$ while the $q\bar{q}$ channel is much smaller.  For the $gg$ and $qg$ channels and when studying the fractional deviations from the effective theory result, we implement the cut $p_T > 15$ GeV, consistent with the CDF and D0 definitions of a jet~\cite{TEVhiggs}.

\subsubsection{Effects of finite fermion masses}

\begin{figure}[htbp]
   \centering
   \includegraphics[width=0.37\textwidth,angle=90]{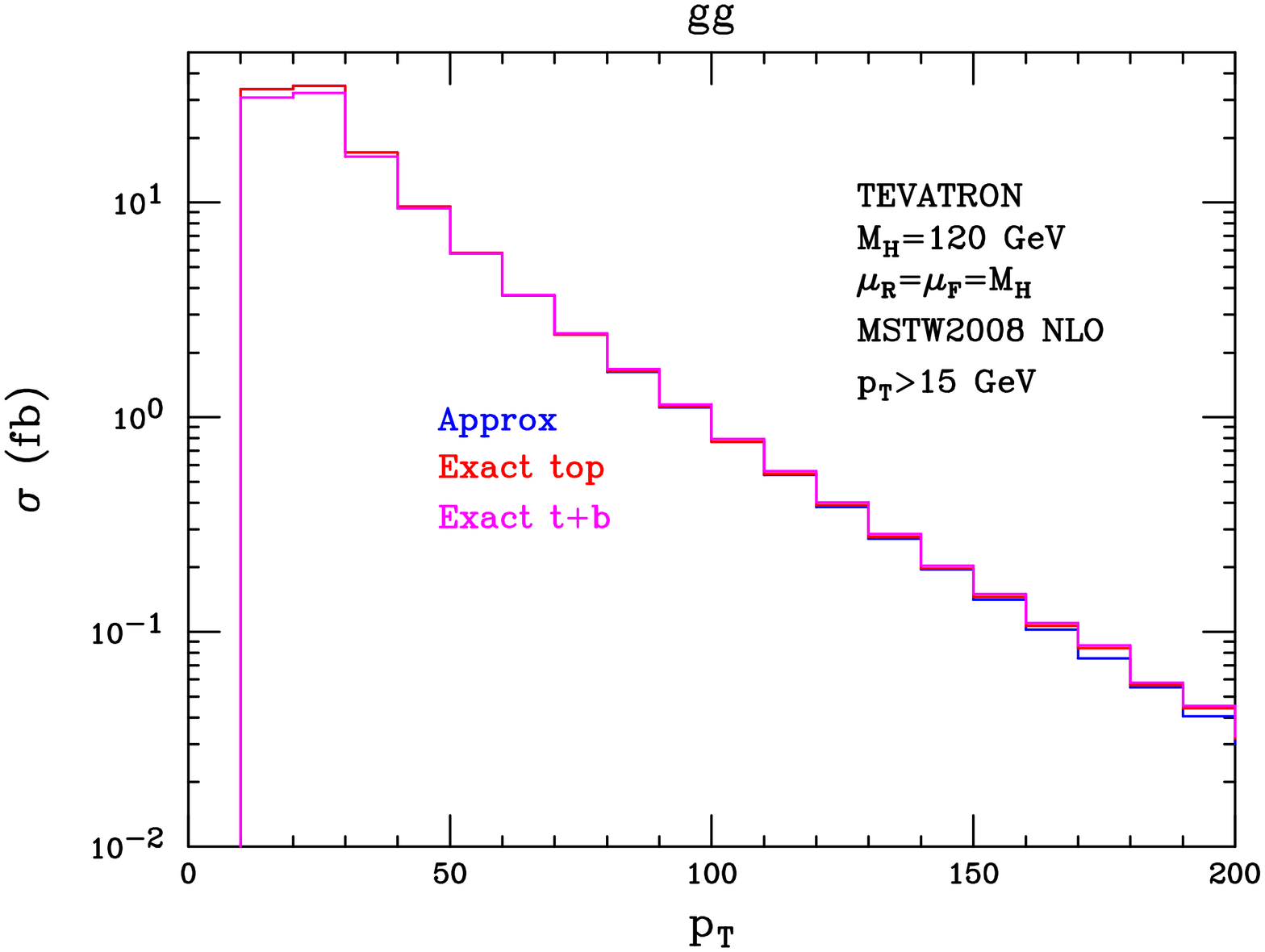}
   \includegraphics[width=0.37\textwidth,angle=90]{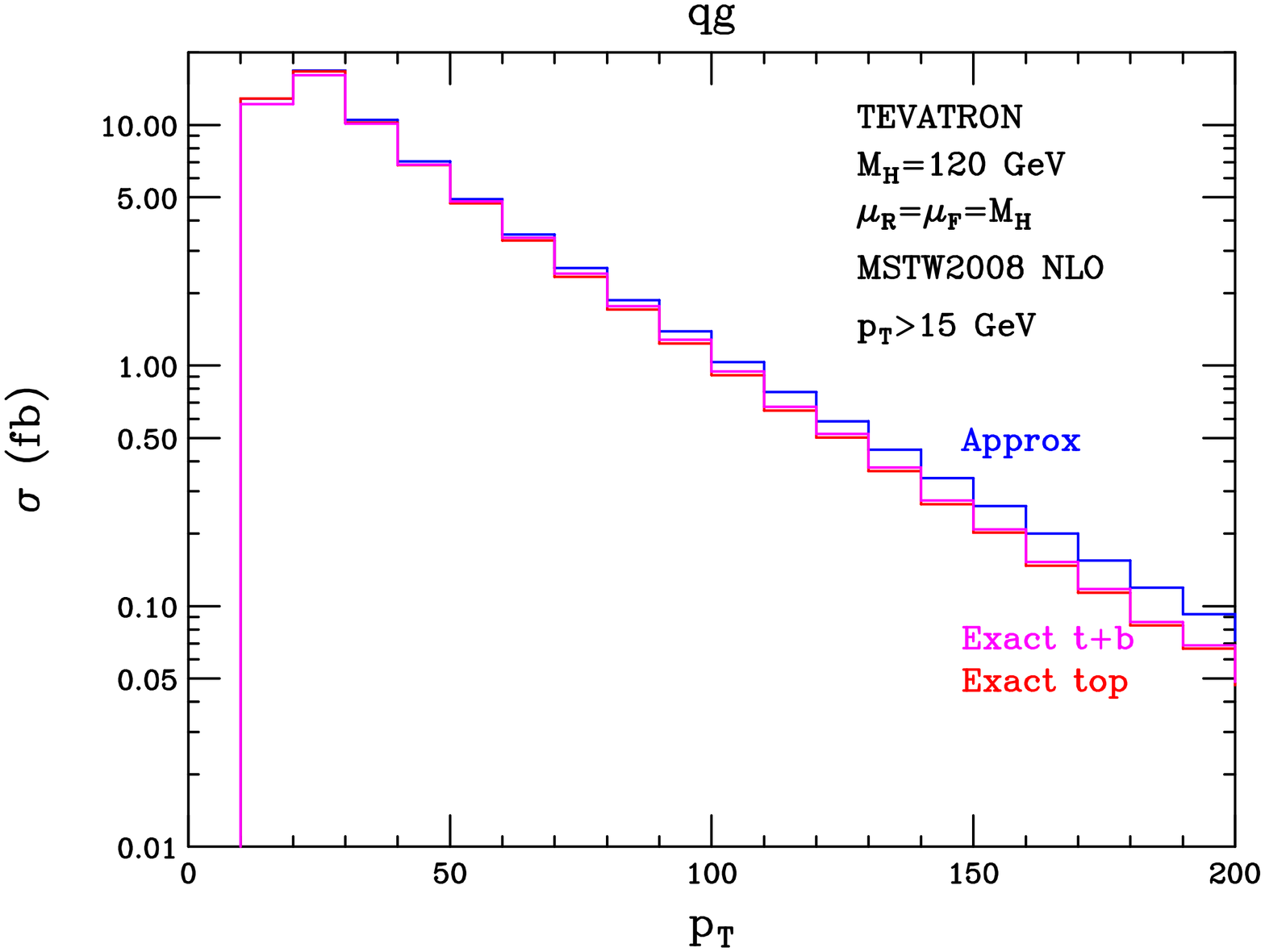}
   \caption{Transverse momentum spectra for the $gg$ (left panel) and $qg$ (right panel) channels at the Tevatron.  A cut $p_T > 15$ GeV is imposed, and a bin size of 10 GeV is 
   used.}
   \label{TeVpt1}
\end{figure}

\begin{figure}[htbp]
   \centering
   \includegraphics[width=0.37\textwidth,angle=90]{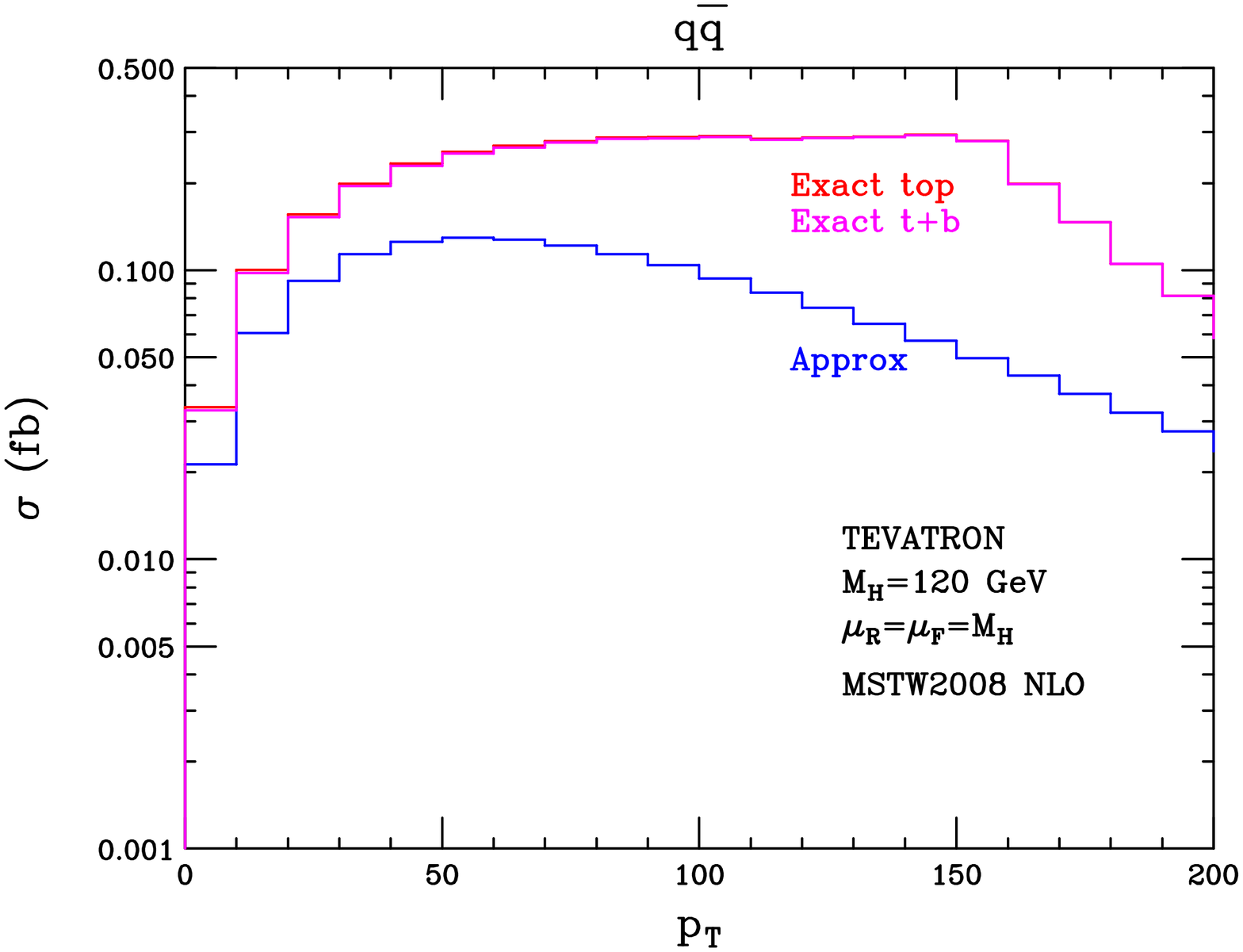}
   \includegraphics[width=0.37\textwidth,angle=90]{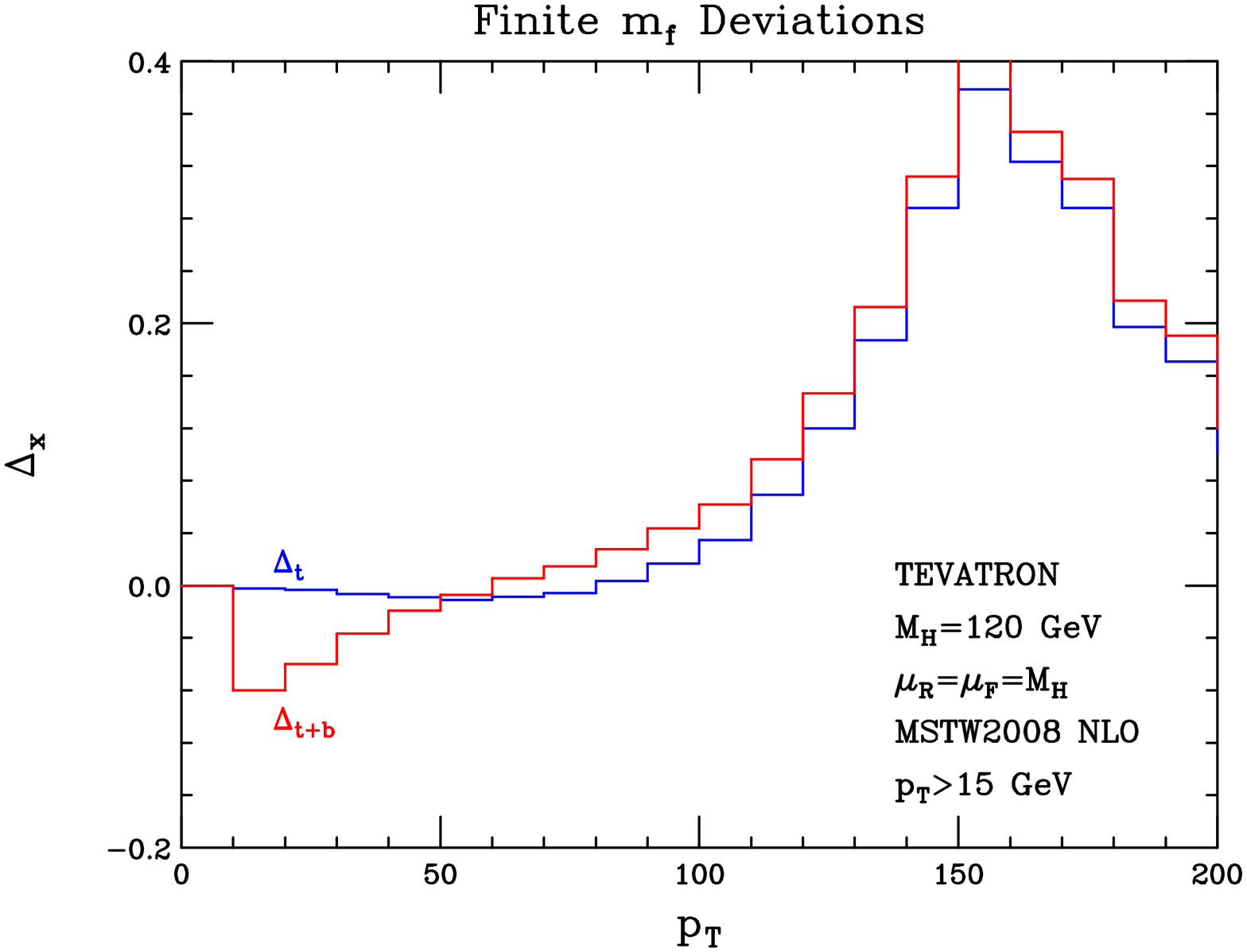}
   \caption{Transverse momentum spectrum for the $q\bar{q}$ partonic channel at the Tevatron (left panel).  A bin size of 10 GeV is used.  The right panel 
   shows the deviations from the approximate cross section for the top quark with its exact mass dependence and for the top and bottom quarks.}
   \label{TeVpt2}
\end{figure}

A typical approximate method for computing this cross section involves integrating out the top quark in the limit $2 m_t \gg M_H$ to produce the effective $Hgg$ 
operator
\begin{equation}
{\cal L}_{EFT} =  -\alpha_s\frac{C_1}{4v} H G_{\mu\nu}^a G^{a\mu\nu}.
\end{equation}
The cross section is computed using this Lagrangian and then normalized to the exact $m_t$-dependent $gg \to H$ interaction, leading to the approximate 
result
\begin{equation}
\sigma^{approx} = \left(\frac{\sigma_{gg\to H}^{exact}}{\sigma_{gg\to H}^{EFT}}\right) \sigma^{EFT}.
\label{cr_approx}
\end{equation}
We begin by quantifying how well this approximation works as a function of Higgs $p_T$.  We first show in Figs.~\ref{TeVpt1} and~\ref{TeVpt2} the $p_T$ spectra for each partonic channel separately for three 
cases: the approximate cross section including just the top quark defined in Eq.~(\ref{cr_approx}); the result with the exact top-quark mass dependence; and the 
cross section including both the top and bottom quarks with their exact mass dependence.  These are shown to display the size of the cross section for various initial states and in each $p_T$ bin.  The sizes of the shifts are not easily seen on a logarithmic scale.  We also define the deviation of the exact full cross section from the approximate form:
\begin{equation}
\Delta_x = \frac{\sigma_x - \sigma^{approx}}{\sigma^{approx}},
\end{equation}
where $x$ indicates either the top-quark result or the cross section including both top and bottom quarks.  Because the gluon parton distribution function rises 
rapidly at small Bjorken-$x$, most events in the $qg$ and $gg$ channels occur with small Mandelstam invariants.  This implies that the loop integrals can be expanded 
around large $m_t$, and the effective theory approximation works well.  This is not true for the $q\bar{q}$ channel, where events with large $\hat{s}$ are not 
highly suppressed.  The deviations from the approximate result are shown in the right panel of Fig.~\ref{TeVpt2}.  Bottom-quark loops in the $gg$ and $qg$ channels 
lead to a destructive interference effect of up to $-8\%$ at $p_T < 50$ GeV.  The effect of the bottom quarks has a dependence on the value of $p_T$, varying from 
$-8\%$ at $p_T \approx 15$ GeV to $+4\%$ at $p_T \approx 100$ GeV.  At high $p_T$ above 100 GeV, where the relative size of the 
$q\bar{q}$ channel is increased, the 
large positive deviations caused by this channel reach $+30\%$ or more.  However, the contribution of this kinematic region to the total rate is very small, as is clear from 
the plots.  The kinematic feature near $p_T \approx 150$ GeV in the exact $q\bar{q}$ distribution occurs because the constraint $\sqrt{\hat{s}} \geq \sqrt{p_T^2+M_H^2}+p_T$ 
leads to $\sqrt{\hat{s}} > 2 m_t$ and the onset of an imaginary part in the amplitude at this value.

\subsubsection{Effects of electroweak corrections}

\begin{figure}[htbp]
   \centering
   \includegraphics[width=0.37\textwidth,angle=90]{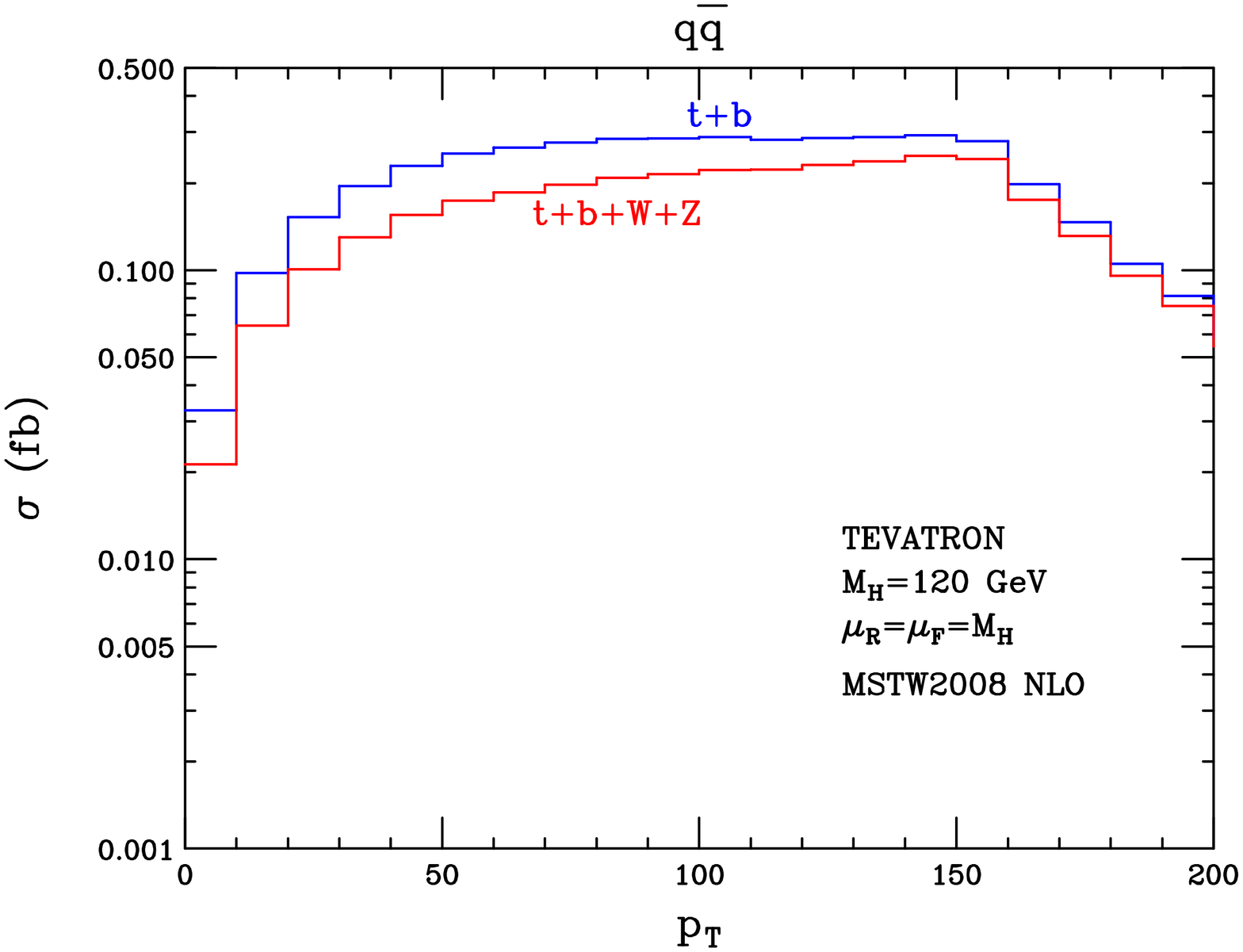}
    \includegraphics[width=0.37\textwidth,angle=90]{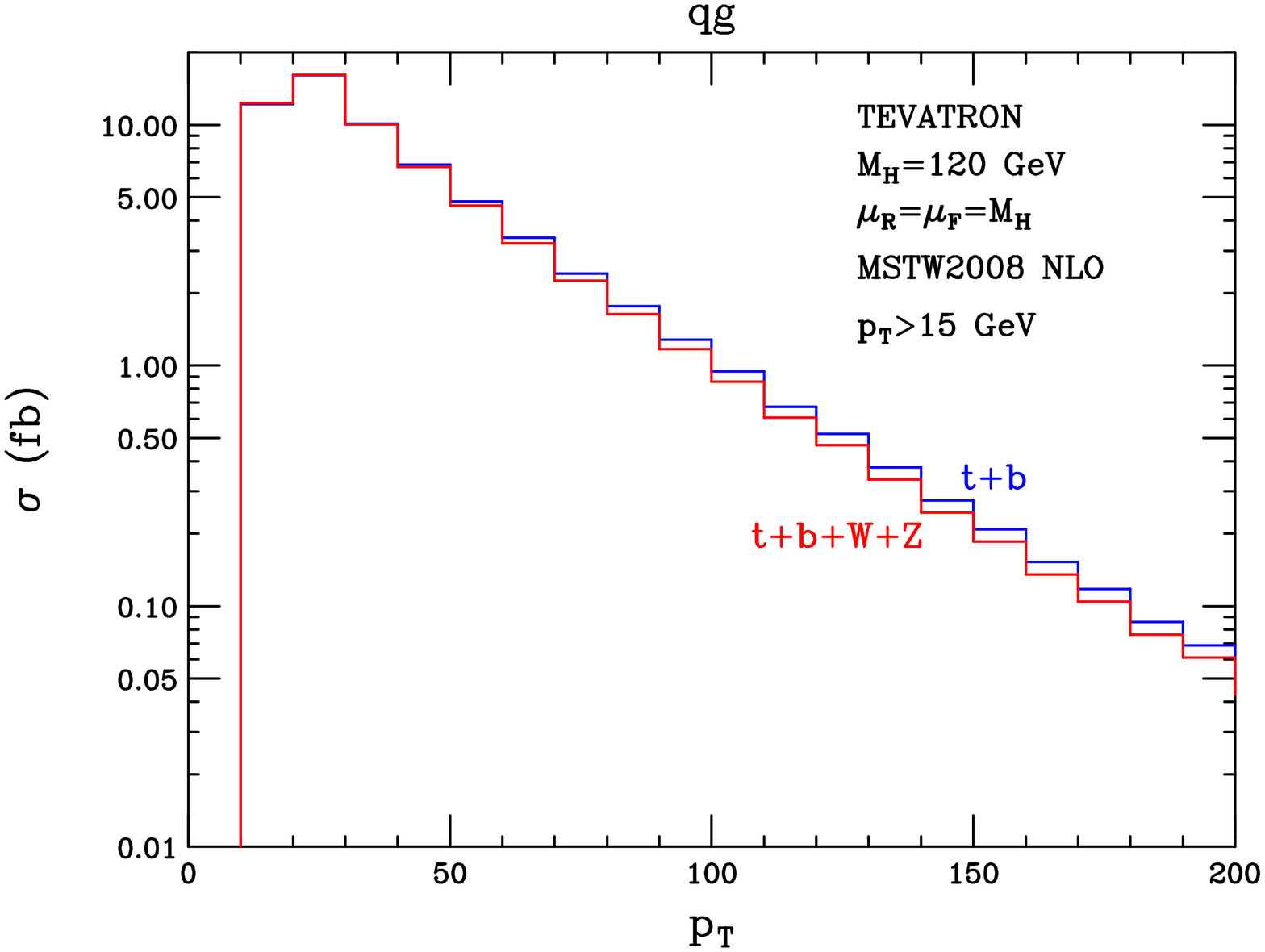}
   \caption{Transverse momentum spectra for the $q\bar{q}$ and $qg$ channels at the Tevatron, with and without the electroweak corrections.  The exact $m_f$-dependent amplitudes have been used 
   to obtain both distributions.}
   \label{EWTeVnums}
\end{figure}

We now study the effects of including the $W$ and $Z$ boson corrections.  These affect only the $qg$ and $q\bar{q}$ partonic channels.  As noted before, these are expected to affect Higgs production predominantly at high $p_T$.  We show in Fig.~\ref{EWTeVnums} the $p_T$ spectra of both the $q\bar{q}$ and $qg$ channels 
for the cases when the top and bottom quarks are included with their exact mass dependence, and when the $W$ and $Z$ are added.  Although the effects on the $q\bar{q}$ channel are quite large, this channel is highly suppressed because of its lower luminosity.  The addition of the $W$ and $Z$-mediated terms 
contributes an additional destructive interference at high transverse momentum in the $qg$ channel.

\begin{figure}[htbp]
   \centering
   \includegraphics[width=0.37\textwidth,angle=90]{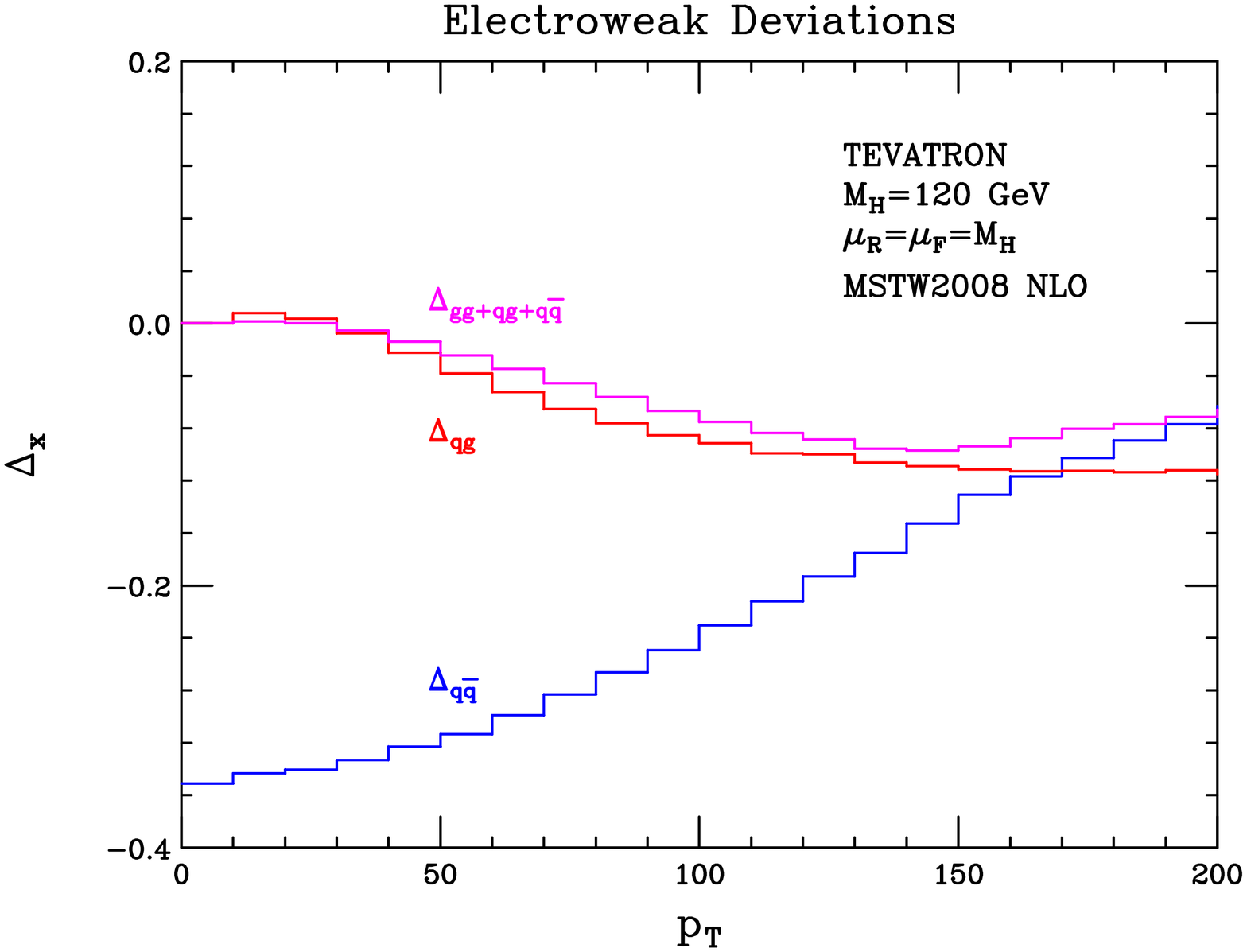}
   \includegraphics[width=0.37\textwidth,angle=90]{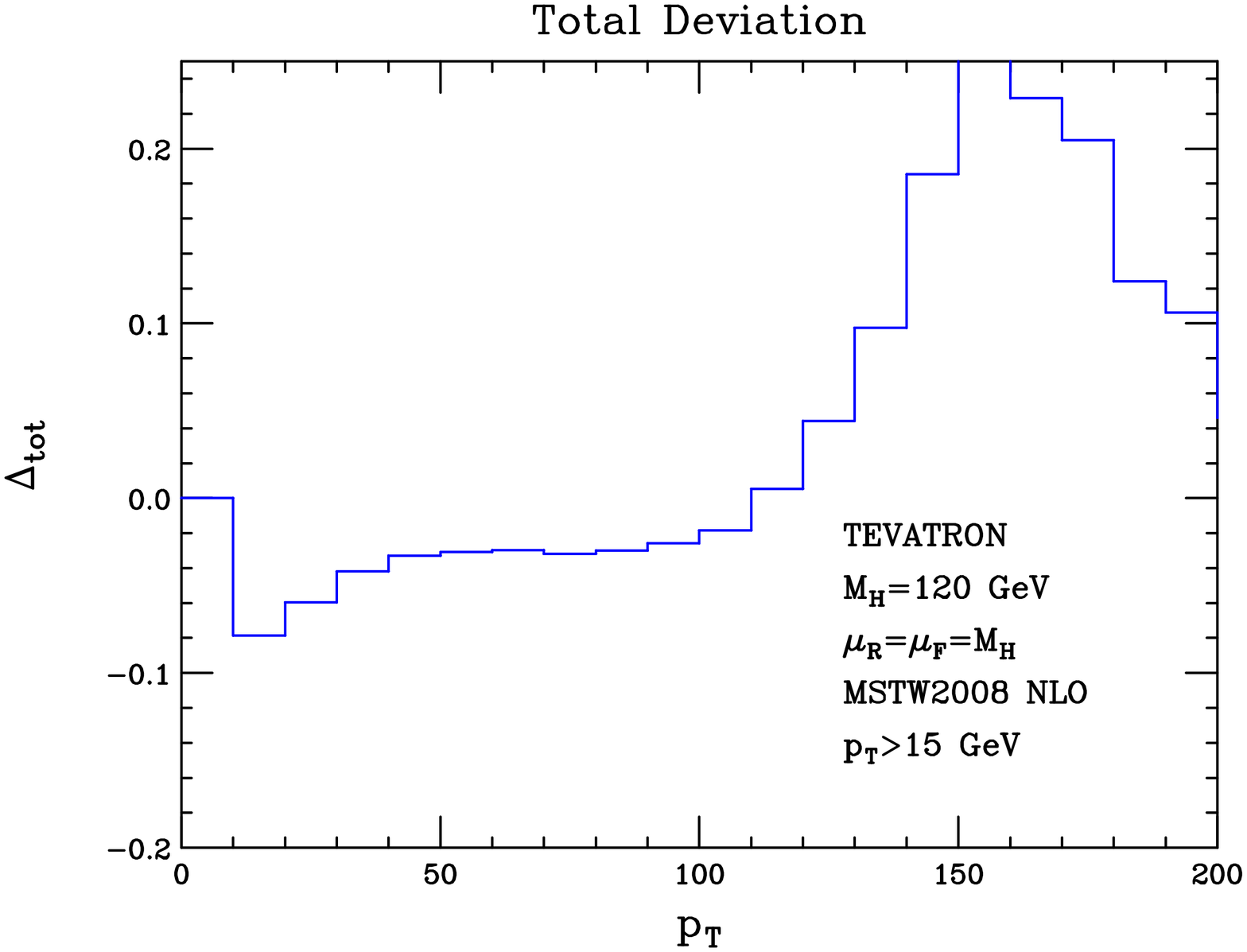}
   \caption{Fractional deviations of the bin-integrated cross sections including the $W$ and $Z$ from that using the top and bottom quarks with their exact mass 
   dependence (left panel) at the Tevatron.  Results are shown for the $qg$ and $q\bar{q}$ channels, and for the total cross section.  Also shown is the total deviation from the 
   approximate top quark result defined in Eq.~(\ref{cr_approx}) arising from $W$, $Z$, $b$ loops and the exact top-quark mass dependence (right panel).}
   \label{TeVdev}
\end{figure}

The fractional deviations in each of the affected partonic channels and in the total cross section are shown in Fig.~\ref{TeVdev}.  The deviations are measured from the cross sections including both the top and bottom-quark terms with their exact mass dependence.  Although the $gg$ partonic channel is the largest by a factor of a few, the electroweak effects are large enough to result in a significant destructive interference for high Higgs $p_T$.  The effects reach $-4\%$ for $p_T \approx 50$ GeV and 
$-8\%$ for $p_T \approx 100$ GeV.  Also shown is the total deviation form the approximate top-quark result defined in Eq.~(\ref{cr_approx}) arising from $W$, $Z$, $b$ loops and the exact top-quark mass dependence.  The total deviation reaches $-8\%$ at $p_T \approx 15$ GeV and remains at $-3\%$ from $p_T \approx 50-100$ GeV.

\subsection{LHC}

We now repeat our numerical study for the LHC, assuming $\sqrt{s} = 10$ TeV.  To avoid singularities and the effects of large logarithms, we implement the cut $p_T > 30$ GeV 
in the $qg$ and $gg$ partonic channels.

\subsubsection{Effects of finite fermion masses}

\begin{figure}[htbp]
   \centering
   \includegraphics[width=0.37\textwidth,angle=90]{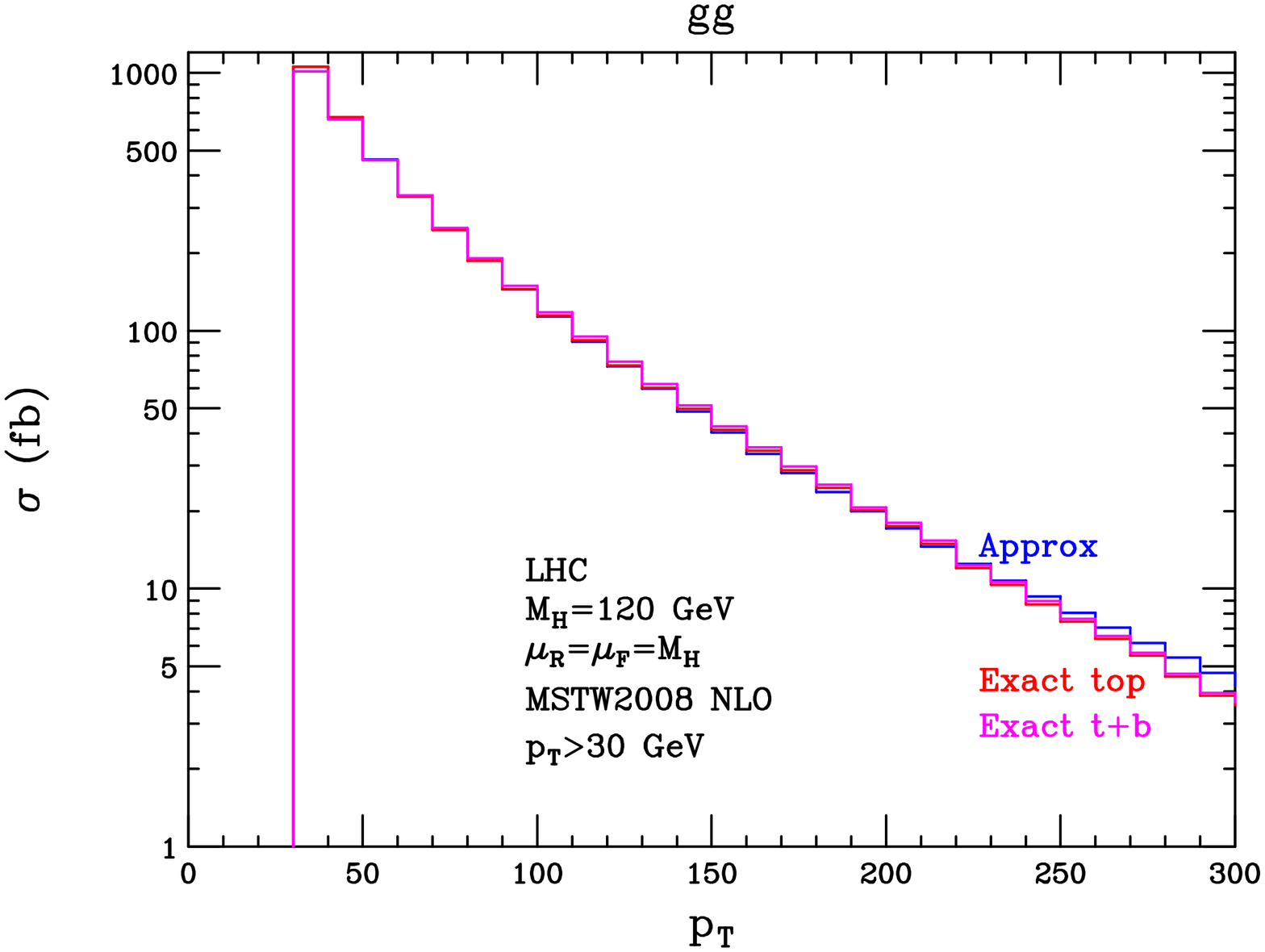}
   \includegraphics[width=0.37\textwidth,angle=90]{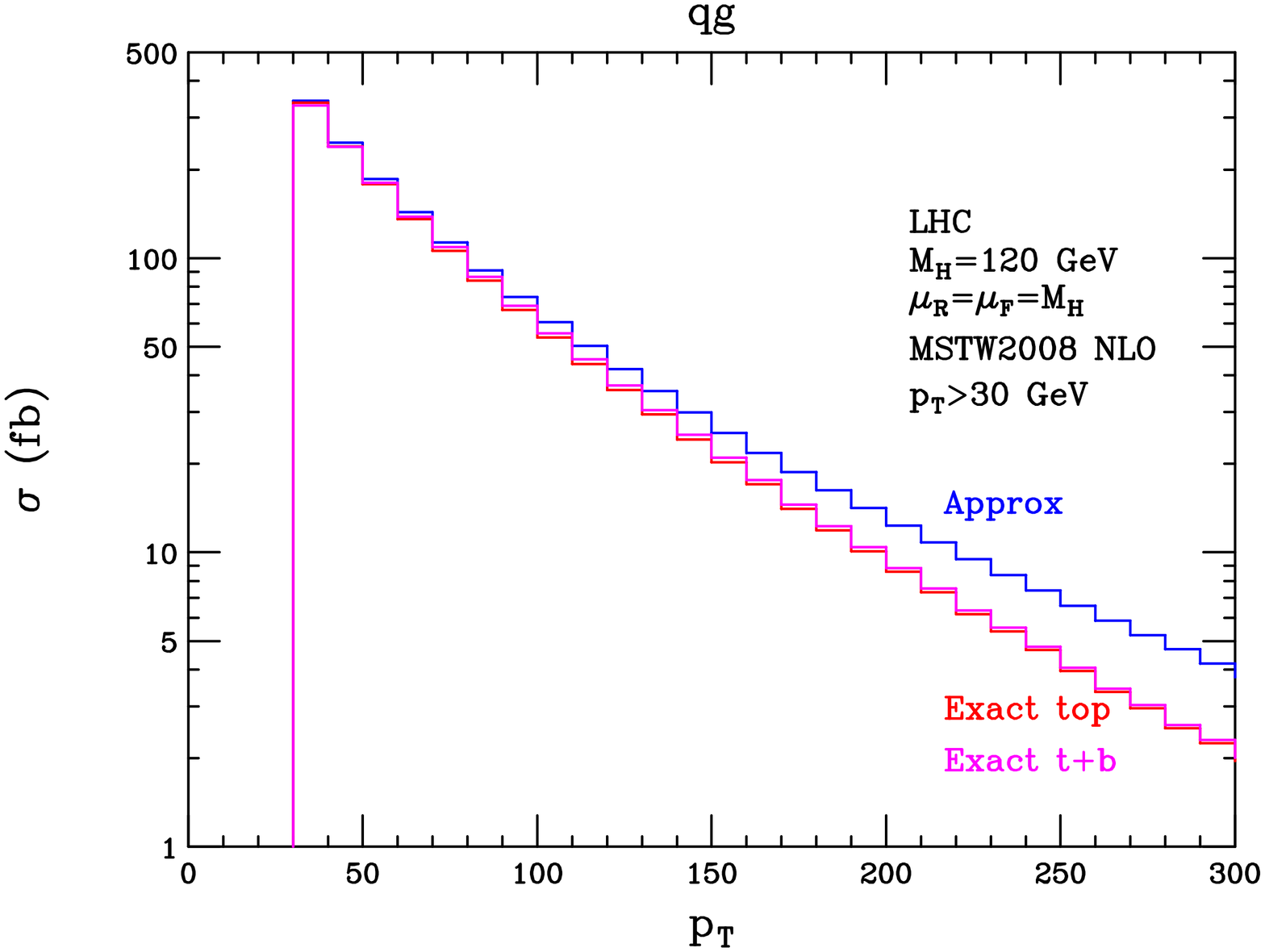}
   \caption{Transverse momentum spectra for the $gg$ (left panel) and $qg$ (right panel) channels at the LHC.  A cut $p_T > 30$ GeV is imposed, and a bin size of 10 GeV is 
   used.}
   \label{LHCpt1}
\end{figure}

\begin{figure}[htbp]
   \centering
   \includegraphics[width=0.37\textwidth,angle=90]{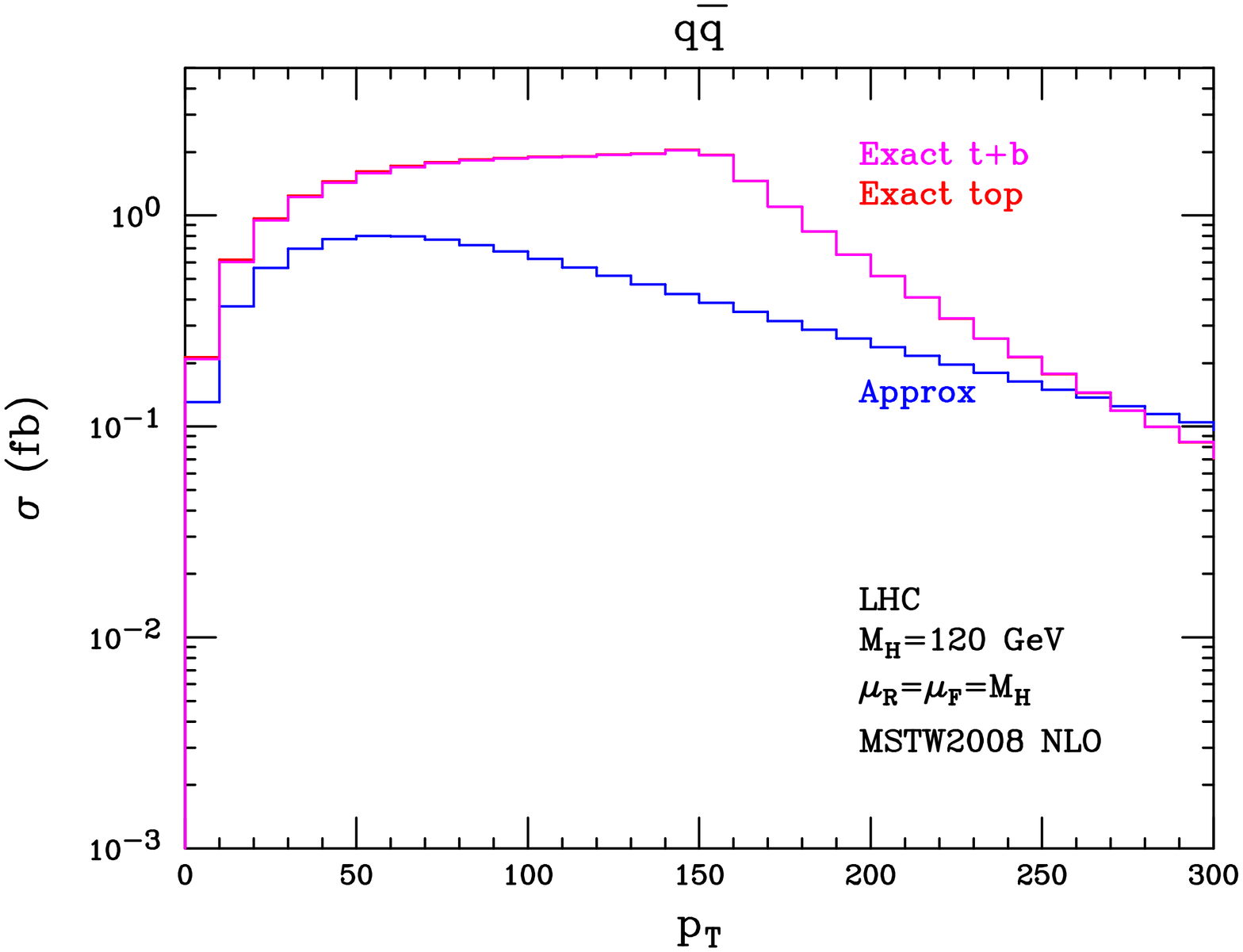}
   \includegraphics[width=0.37\textwidth,angle=90]{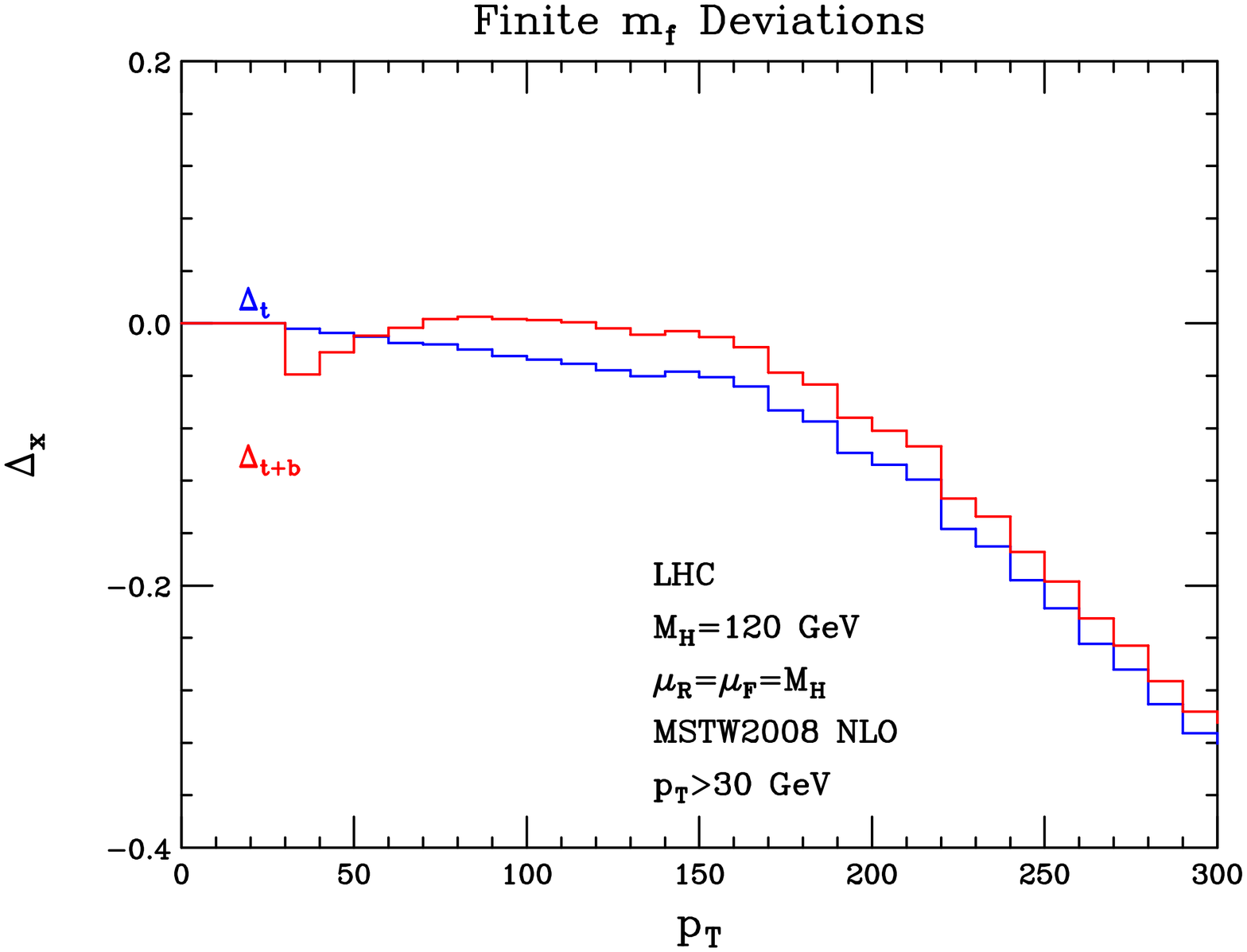}
   \caption{Transverse momentum spectrum for the $q\bar{q}$ partonic channel at the LHC (left panel).  A bin size of 10 GeV is used.  The right panel 
   shows the deviations from the approximate cross section for the top quark with its exact mass dependence and for the top and bottom quarks.}
   \label{LHCpt2}
\end{figure}

We again begin by studying the effect of using the approximate cross section defined in Eq.~(\ref{cr_approx}) versus the exact result using the full top and bottom quark mass dependence.  The transverse momentum spectra for the various partonic channels are again shown in Figs.~\ref{LHCpt1} and~\ref{LHCpt2} for orientation.  One important difference from the Tevatron case is that the large 
deviations in the $q\bar{q}$ channel do not become visible over the $gg$ and $qg$ channels even for transverse momenta of a few hundred GeV.  The dominant effects up to $p_T=300$ GeV are therefore 
the deviations caused by using the exact $m_t$-dependent amplitudes and including bottom quarks in the gluon-initiated channels.  These are shown in Fig.~\ref{LHCpt2}; they can be 
quite large and have a complicated dependence on $p_T$ due to cancellation between the shifts caused by the top and bottom-quark terms.  The deviations resulting from including both the 
exact top and bottom-quark mass dependence are $-4\%$ for $30\,{\rm GeV} \leq p_T \leq 50\,{\rm GeV}$, 
at the percent-level or less from 50 GeV until 150 GeV, and then become increasingly negative, reaching $-30\%$ at $p_T \approx 300$ GeV.

\subsubsection{Effects of electroweak corrections}

\begin{figure}[htbp]
   \centering
   \includegraphics[width=0.35\textwidth,angle=90]{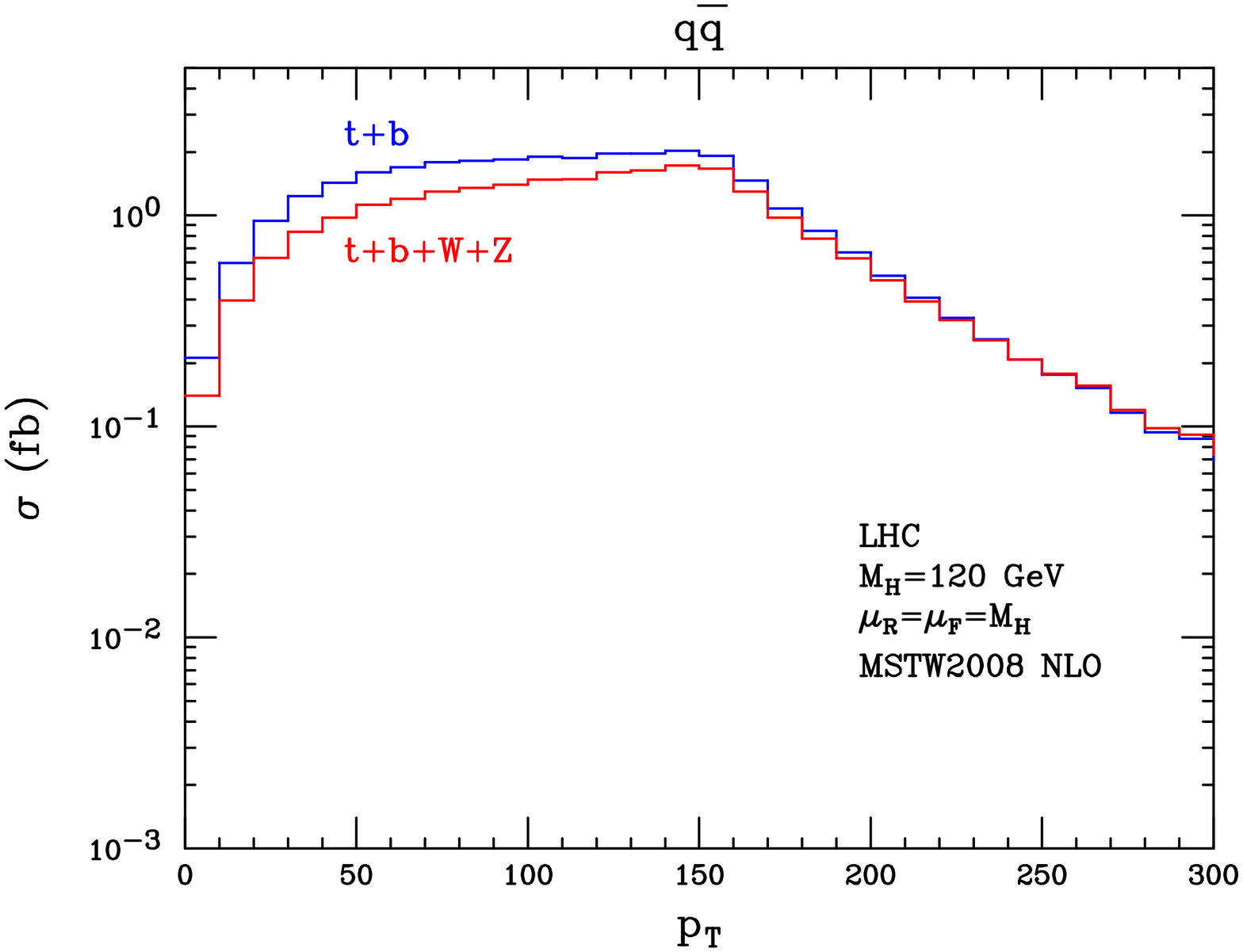}
    \includegraphics[width=0.35\textwidth,angle=90]{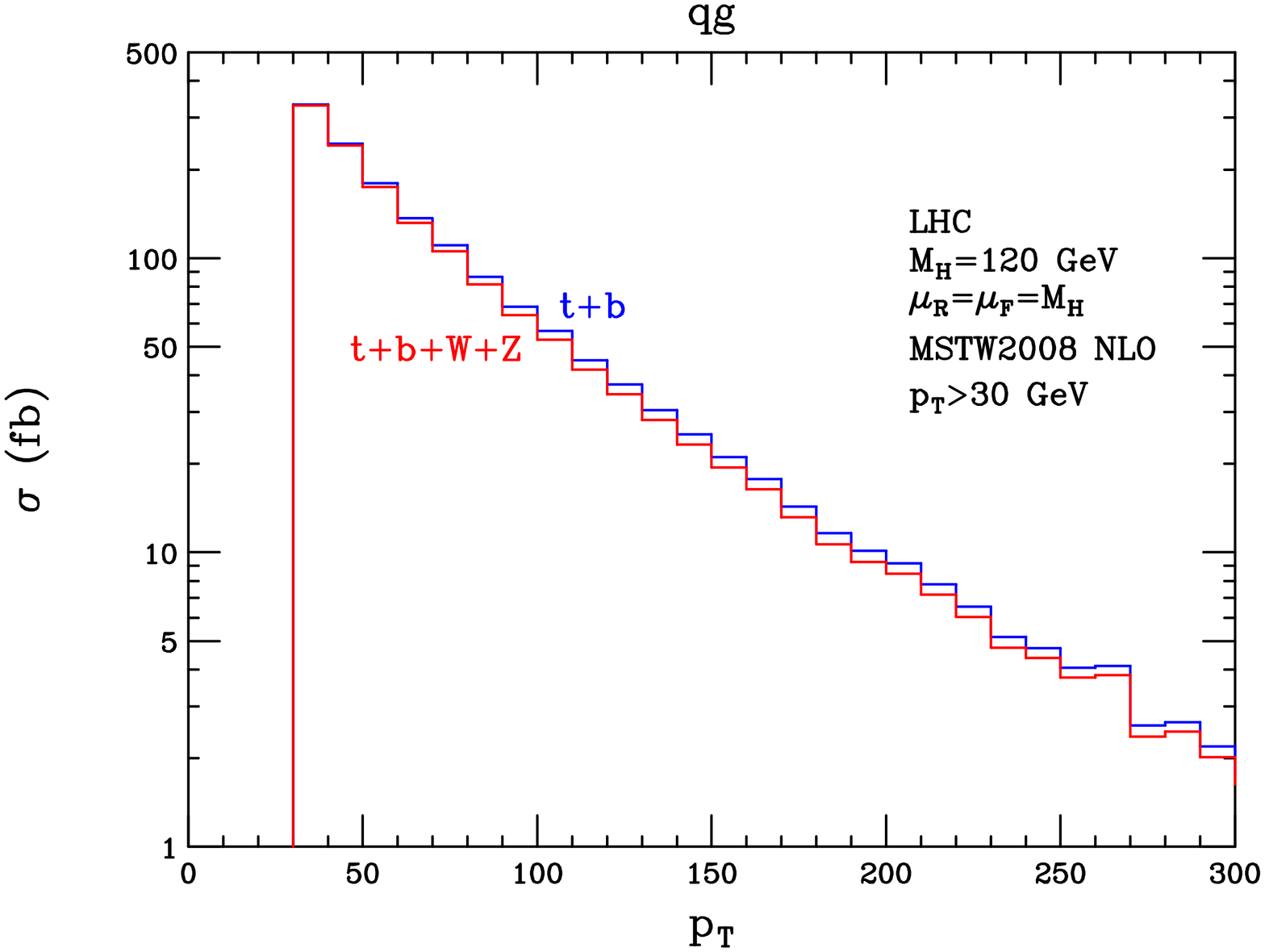}
   \caption{Transverse momentum spectra for the $q\bar{q}$ and $qg$ channels at the LHC, with and without the electroweak corrections.  The exact $m_f$-dependent amplitudes have been used 
   to obtain both distributions.}
   \label{EWLHCnums}
\end{figure}

We now study the effect of adding $W$ and $Z$-induced electroweak corrections at the LHC.  We expect them to be smaller than at the Tevatron, as the gluon-initiated partonic channels dominate to a larger extent at the LHC.  The transverse momentum spectrum for the $q\bar{q}$ and $qg$ channels, both with and without the electroweak terms, are shown in Fig.~\ref{EWLHCnums}.  Again, the deviations in the $q\bar{q}$ channel are large, but the numerical impact 
of this component is suppressed by its low luminosity.

\begin{figure}[htbp]
   \centering
   \includegraphics[width=0.37\textwidth,angle=90]{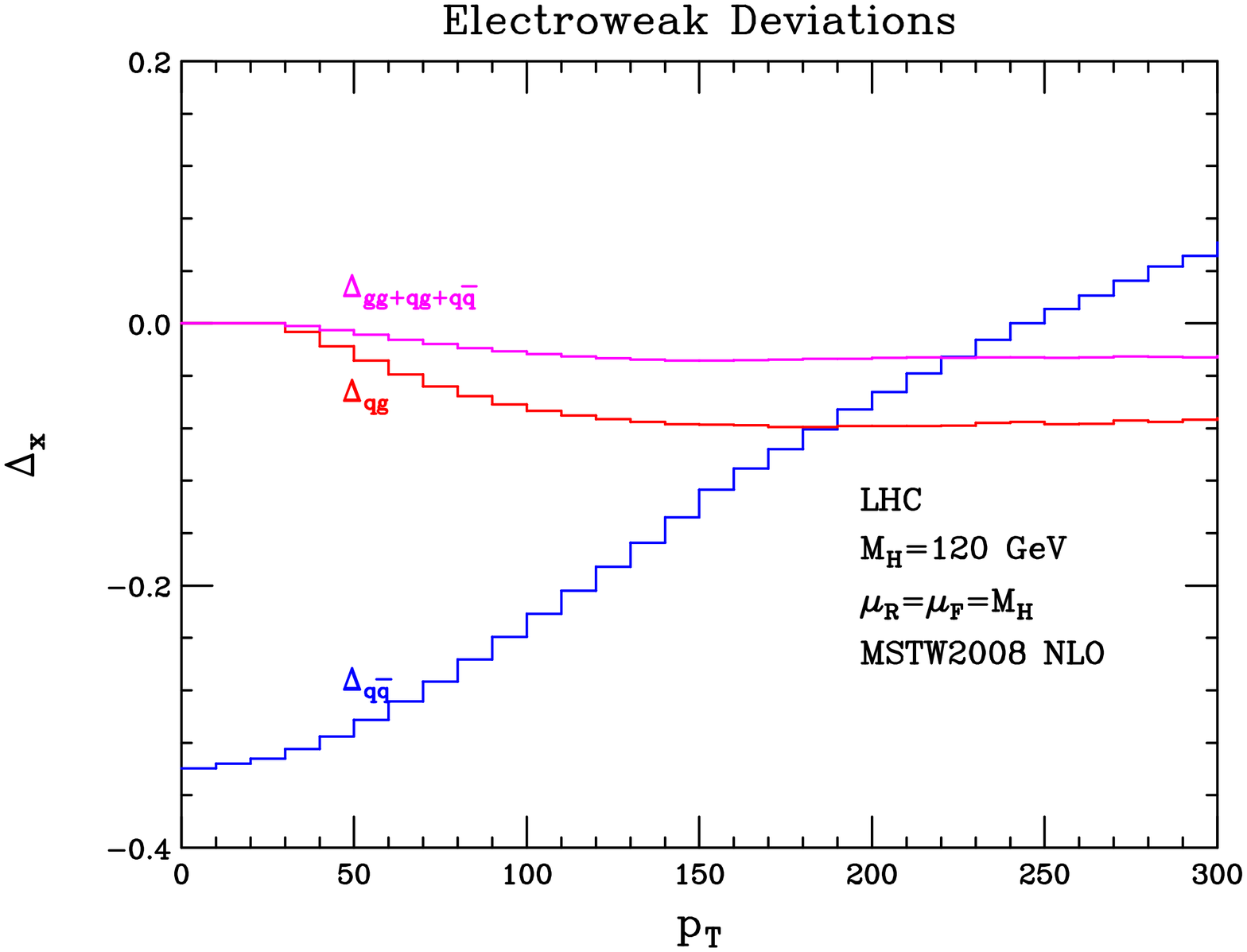}
   \includegraphics[width=0.37\textwidth,angle=90]{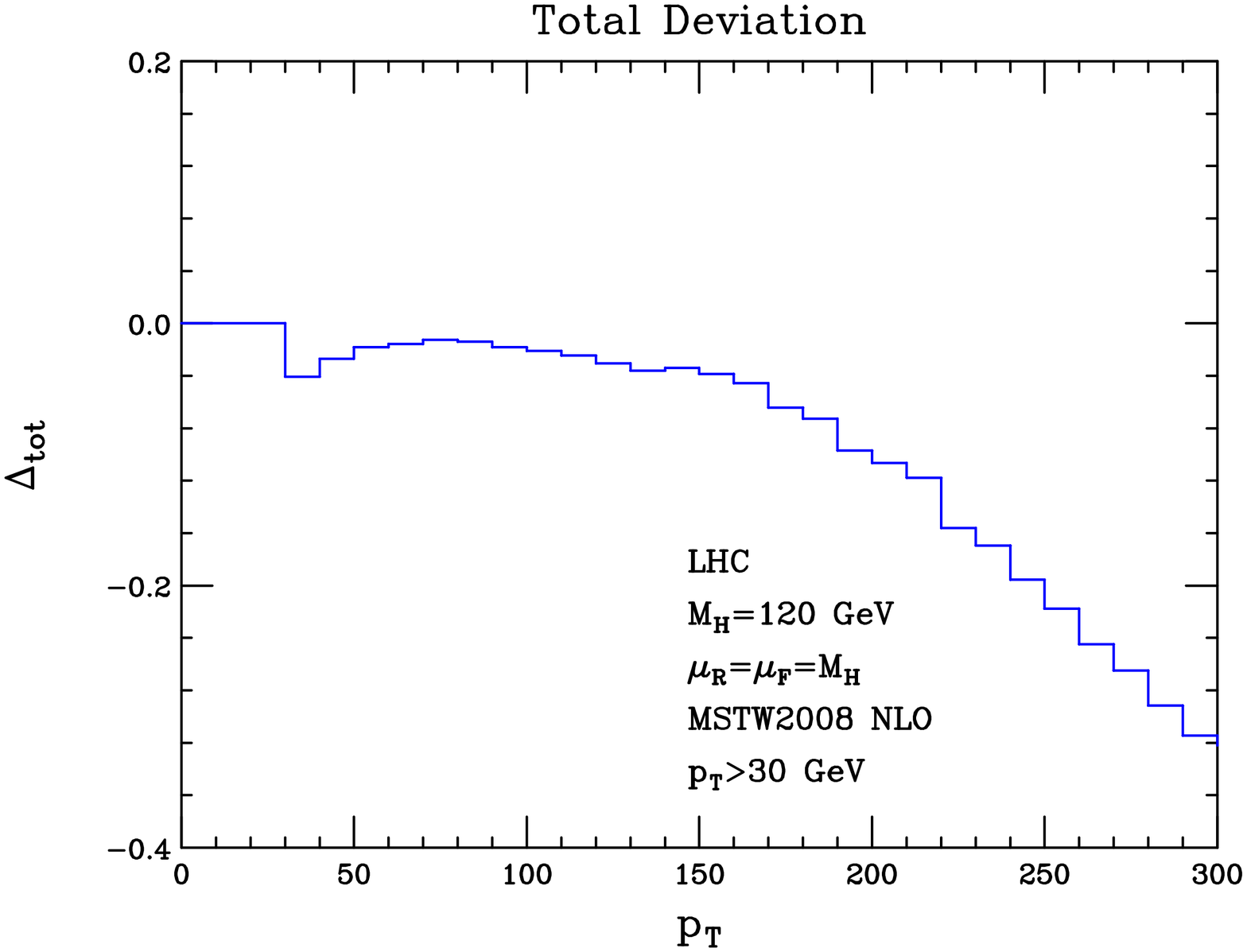}
   \caption{Fractional deviations of the bin-integrated cross sections including the $W$ and $Z$ from that using the top and bottom quarks with their exact mass 
   dependence (left panel) at the LHC.  Results are shown for the $qg$ and $q\bar{q}$ channels, and for the total cross section.  Also shown is the total deviation from the 
   approximate top quark result defined in Eq.~(\ref{cr_approx}) arising from $W$, $Z$, $b$ loops and the exact top-quark mass dependence (right panel).}
   \label{LHCdev}
\end{figure}

The fractional deviations for each partonic channel and for the total cross section caused by the electroweak effects are shown in Fig.~\ref{LHCdev}.  As expected, the results 
are smaller than at the Tevatron.  The $W$ and $Z$-mediated contributions induce a $-3\%$ shift for $p_T$ values between 100 and 300 GeV.  Also shown is the total deviation form the approximate top-quark result defined in Eq.~(\ref{cr_approx}) arising from $W$, $Z$, $b$ loops and the exact top-quark mass dependence.  This shift shrinks from $-4\%$ at $p_T \approx 30$ GeV 
to $-1\%$ at $p_T \approx 100$ GeV, and then continuously grows to reach $-32\%$ at $p_T=300$ GeV.

\section{Conclusions}
\label{sec:conc}

In this paper we have performed a detailed study of the possible one-loop contributions to the Higgs boson transverse momentum spectrum.  We have considered effects arising from bottom quarks and the $W,Z$ electroweak gauge bosons, as well as corrections arising from the full mass dependence of the top quark.  The deviations from the often-used approximation defined in 
Eq.~(\ref{cr_approx}), where only an infinitely heavy top quark is included, are generically negative, reaching $-8\%$ for moderate $p_T$ at the Tevatron and $-30\%$ for transverse momenta of several hundred GeV at the LHC.  The corrections have a non-neglible dependence on the Higgs $p_T$.  At the LHC, they may be particularly important if regions of large transverse momenta are selected, which has been discussed to increase to discovery potential in several Higgs boson search channels~\cite{Abdullin:1998er,Mellado:2007fb}.  The electroweak corrections reach $-8\%$ at the Tevatron, and could potentially have an effect on the current exclusion limits.   As far as we are aware, these contributions have not been implemented in any available calculation, nor in any experimental analysis.  If a scalar is observed at the LHC, these effects should be accounted for to avoid mistakenly claiming a hint of new physics. 

\newpage
\bigskip
\noindent
{\Large\bf Acknowledgements}

\bigskip
\noindent
We thank M. Herndon for helpful discussions.  The work of W.-Y. K. is supported by the DOE grant DE-FG02-84ER40173.  The work of F. P. is supported by the DOE grant DE-FG02-95ER40896, Outstanding Junior Investigator Award and
by the Alfred P.~Sloan Foundation.


\section*{Appendix}
\setcounter{section}{5}

We present here the form factors that appear in the various partonic processes.  We begin with the form factors ${\cal F}^x(\hat{t},\hat{u},\hat{s},M_H,m_x)$ defined in 
Eq.~(\ref{amp}) that compose the $q\bar{q} \to Hg$ and $qg \to Hq$ channels.  We recall that the Mandelstam invariants satisfy $\hat{s}+\hat{t}+\hat{u}=M_H^2$.  For the fermionic contributions, where $x=t,b$, we find
\begin{eqnarray}
{\cal F}^f(\hat{t},\hat{u},\hat{s},M_H,m_f) &=& \frac{8\, m_f}{\hat{s}-M_H^2} \left\{\frac{{\rm Bub1}(M_H^2,m_f^2)-{\rm Bub1}(\hat{s},m_f^2)}{\hat{s}-M_H^2}\right. 
\nonumber \\ &&+\frac{(M_H^2-s-4\, m_f^2)\,{\rm TriF}(\hat{s},M_H^2,m_f^2)}{2\,\hat{s}}  \left. +\frac{1}{\hat{s}}\right\}.
\end{eqnarray}
For the electroweak contributions with $x=W,Z$, we find
\begin{eqnarray}
{\cal F}^V(\hat{t},\hat{u},\hat{s},M_H,M_V) &=&
2\,{\frac { \left( -{\it \hat{t}}+{\it M_V^2} \right)  \left( -{\it \hat{t}}\,{\it \hat{u}}+{\it M_V^2}\,{\it \hat{u}}+{\it M_V^2}\,{\it \hat{t}} \right) {\rm Box}(\hat{t},\hat{u},M_H^2,M_V^2)}{{\it \hat{s}}\,{{\it \hat{t}}}
^{2}}}
\nonumber \\ &&\hspace{-4.0cm}
-2{\frac {{\rm Bub2} \left( {\it \hat{t}},M_V^2 \right) }{{\it \hat{s}}\,{\it \hat{t}}}}+2
\,{\frac { \left( {\it \hat{u}}+{\it \hat{s}} \right)  \left( -{\it \hat{t}}+{\it M_V^2} \right) {\rm Tri} \left( {\it \hat{t}},M_H^2,M_V^2 \right) }{{\it \hat{s}}\,{{\it \hat{t}}}^{2}}}
+4\,{\frac {{\rm Bub1}(M_H^2,M_V^2)}{{\it \hat{t}}\, \left( {\it \hat{t}}+{\it \hat{s}} \right) }}
\nonumber \\ &&\hspace{-4.0cm}
-2\,{\frac {
 \left( -{{\it \hat{s}}}^{2}{\it \hat{t}}+{{\it \hat{s}}}^{2}{\it M_V^2}+2\,{\it M_V^2}\,{\it \hat{s}}\,{\it \hat{t}}-{\it M_V^2}\,{{\it \hat{t}}}^{2}+{{\it \hat{t}}}^{3} \right) {\rm Tri}
 \left( {\it \hat{u}},M_H^2,M_V^2 \right) }{ \left( {\it \hat{t}}+{\it \hat{s}} \right) {\it \hat{s}}\,{{\it \hat{t}}}^{2}}}
  \nonumber \\ &&\hspace{-4.0cm}
  -2{
\frac { \left( -{\it \hat{t}}+{\it M_V^2} \right) {\it \hat{u}}\,{\rm Bub2} \left( {\it \hat{u}},M_V^2 \right) }{ \left( -{\it \hat{u}}+{\it M_V^2} \right) {\it \hat{s}}\,{{\it \hat{t}}}^{2}}} 
 +6\,\frac{ \left( -{\it \hat{t}}+{\it M_V^2} \right) \ln  \left( 1-{\frac {{\it 
\hat{t}}}{{\it M_V^2}}} \right)} {\hat{s} \,\hat{t}^2}
\nonumber \\ &&\hspace{-4.0cm}
+2\, \left( -3\,{{\it \hat{t}}}^{2}{\it \hat{u}}-5\,{\it \hat{t}}\,{\it \hat{u}}\,{\it \hat{s}}+2\,{\it M_V^2}\,{\it \hat{s}}\,{
\it \hat{t}}+3\,{\it \hat{s}}\,{\it \hat{u}}\,{\it M_V^2}+3\,{\it M_V^2}\,{\it \hat{t}}\,{\it \hat{u}} \right) \ln  \left( 1-{\frac {{\it \hat{u}}}{{\it M_V^2}}} \right)  \left( {\it \hat{t}}+{
\it \hat{s}} \right) ^{-1} 
\nonumber \\ && \hspace{-4.0cm}
\times {{\it \hat{s}}}^{-1}{{\it \hat{t}}}^{-2}{{\it \hat{u}}}^{-1}+8\,{\frac {2\,{\it M_V^2}\,{\it \hat{s}}\,{\it \hat{t}}+{\it M_V^2}\,{\it \hat{t}}\,{\it \hat{u}}-2\,{{\it \hat{t}}}
^{2}{\it \hat{u}}+{\it \hat{s}}\,{\it \hat{u}}\,{\it M_V^2}-3\,{\it \hat{t}}\,{\it \hat{u}}\,{\it \hat{s}}+{\it M_V^2}\,{{\it \hat{t}}}^{2}}{ \left( -{\it \hat{u}}+{\it M_V^2} \right) {\it \hat{s}}\,{{\it 
\hat{t}}}^{2} \left( {\it \hat{t}}+{\it \hat{s}} \right) }}.
\end{eqnarray}
The various auxiliary functions that appear in this expression are the finite parts of loop integrals that appear in the calculation; all poles that appear in each integral 
cancel in the amplitude.  They are given by finite parts of the following expressions:
\begin{eqnarray}
{\rm Bub1}(p^2,m^2) &=& \left[ \int \frac{d^d k}{i \pi^{d/2}} \frac{1}{k^2-m^2} \frac{1}{(k+p)^2-m^2} \right]_{finite}, \nonumber \\
{\rm Bub2}(p^2,m^2) &=& \left[ \int \frac{d^d k}{i \pi^{d/2}} \frac{1}{k^2} \frac{1}{(k+p)^2-m^2} \right]_{finite},\nonumber \\
{\rm Tri}(p_1^2,p_2^2,m^2) &=& \left[ \int \frac{d^d k}{i \pi^{d/2}} \frac{1}{k^2} \frac{1}{(k+p_1)^2-m^2} \frac{1}{(k+p_1+p_2)^2-m^2} \right]_{finite} \nonumber \\ 
 && {\rm with} \,\, (p_1+p_2)^2=0, \nonumber \\ 
{\rm TriF}(p_1^2,p_2^2,m^2) &=& \int \frac{d^d k}{i \pi^{d/2}} \frac{1}{k^2-m^2} \frac{1}{(k+p_1)^2-m^2} \frac{1}{(k+p_1+p_2)^2-m^2} \nonumber \\ 
 && {\rm with} \,\, (p_1+p_2)^2=0, \nonumber \\ 
{\rm Box}(p_1^2,p_2^2,p_3^2,m^2) &=&  \left[ \int \frac{d^d k}{i \pi^{d/2}} \frac{1}{k^2} \frac{1}{(k+p_2)^2-m^2} \frac{1}{(k+p_2-p_3)^2-m^2} \right. \nonumber \\ 
&\times& \left. \frac{1}{(k+p_1+p_2-p_3)^2} \right]_{finite} \,\,\,\, {\rm with} \,\, (p_1-p_3)^2=0,\nonumber \\ && (p_2-p_3)^2=0,\,\,\,\,(p_1+p_2-p_3)^2=0.
\end{eqnarray}
It is straightforward to express these integrals in terms of logarithms and dilogarithms using standard techniques~\cite{scalints}.  All integrals were checked numerically against FF~\cite{vanOldenborgh:1990yc} for a variety of phase-space points.  A further useful check on the exact calculation is comparison with the approximate forms of the amplitude in the limit that the internal loop particles become much heavier than the Higgs boson or the 
Mandelstam invariants.  By direct expansion of the relevant diagrams, we can derive the following simple expansions for the form factors:
\begin{eqnarray}
{\cal F}^f(\hat{t},\hat{u},\hat{s},M_H,m_f) &\stackrel{m_f \longrightarrow \infty}{\to}& -\frac{4}{3\,m_f\, \hat{s}},\nonumber \\
{\cal F}^V(\hat{t},\hat{u},\hat{s},M_H,M_V) &\stackrel{M_V \longrightarrow \infty}{\to}& -\frac{5}{9\,M_V^4}.
\end{eqnarray}
We note that these expanded forms do not provide a suitable numerical approximation of the electroweak effects.

For the ${\cal G}_i$ form factors defined in Eq.~(\ref{ampgg}), we first note that only two are independent due to the following symmetry relations:
\begin{eqnarray}
{\cal G}_1(\hat{t},\hat{u},\hat{s},M_H,m_f) &=& {\cal G}_3(\hat{s},\hat{u},\hat{t},M_H,m_f), \nonumber \\
{\cal G}_1(\hat{t},\hat{u},\hat{s},M_H,m_f) &=& -{\cal G}_2(\hat{u},\hat{s},\hat{t},M_H,m_f),
\end{eqnarray}
which imply that only ${\cal G}_1$ and ${\cal G}_4$ are needed to describe the amplitude.  These take the following forms:
\begin{eqnarray}
{\cal G}_1(\hat{t},\hat{u},\hat{s},M_H,m_f) &=& -{\frac {{\it m_f}\,{\it \hat{t}}\, \left( 8\,{{\it m_f}}^{2}{\it \hat{u}}-{\it \hat{s}}\,{\it \hat{u}}
-{{\it \hat{u}}}^{2}+2\,{\it \hat{t}}\,{\it \hat{s}} \right) {\rm BoxT} \left( {\it \hat{s}},{\it \hat{t}},{\it M_H^2},{{\it m_f}}^{2} \right) }{{{\it \hat{u}}}^{2}}} 
\nonumber \\ && \hspace{-4.0cm}
-{\frac {{\it m_f}\, \left( -{\it \hat{t}}\,{\it \hat{s}}+8\,{{\it m_f}}^{2}{\it \hat{t}}-{{\it \hat{t}}}^{2}+2\,{\it \hat{s}}\,{\it \hat{u}} \right) {
\rm BoxT} \left( {\it \hat{s}},{\it \hat{u}},{\it M_H^2},{{\it m_f}}^{2} \right) }{{\it \hat{t}}}} -4\,{\frac {{\it m_f}\,{\it \hat{t}}\, \left( {\it \hat{s}
}-{\it \hat{u}} \right) {\rm Bub1} \left( {\it \hat{t}},{{\it m_f}}^{2} \right) }{ \left( {\it \hat{u}}+{\it \hat{s}} \right) {\it \hat{s}}\,{\it \hat{u}}}}
\nonumber \\ && \hspace{-4.0cm}
+2\,{\frac {{\it m_f}\,{\it \hat{t}}\, \left( 4\,{{\it m_f}}^{2}{\it \hat{s}}+{\it \hat{t}}\,{\it \hat{u}}
 \right) {\rm BoxT} \left( {\it \hat{t}},{\it \hat{u}},{\it M_H^2},{{\it m_f}}^{2} \right) }{{{\it \hat{s}}}^{2}}}-4\,{\frac { \left( {{\it \hat{u}}}^{2}+{{\it \hat{t}}}^{2}+4\,{\it \hat{t}}\,{\it \hat{u}}
 \right) {\it m_f}\,{\rm Bub1} \left( {\it \hat{s}},{{\it m_f}}^{2} \right) }{ \left( {\it \hat{u}}+{\it \hat{t}} \right) ^{2}{\it \hat{u}}}}
 \nonumber \\ && \hspace{-4.0cm}
-4\,{\frac {{\it m_f}\, \left( {\it \hat{s}}-{
\it \hat{t}} \right) {\rm Bub1} \left( {\it \hat{u}},{{\it m_f}}^{2} \right) }{{\it \hat{s}}\, \left( {\it \hat{t}}+{\it \hat{s}} \right) }}-2\,{\frac {{\it m_f}\,{{\it \hat{t}}}^{2}{\rm TriF2} \left( {\it \hat{t}},{{\it m_f}}^{2} \right)  \left( {{\it \hat{s}}}^{2}-{{
\it \hat{u}}}^{2} \right) }{{{\it \hat{u}}}^{2}{{\it \hat{s}}}^{2}}}-2\,{\frac {{\it m_f}\,{\it \hat{u}}\,{\rm TriF2} \left( {\it \hat{u}},{{\it m_f}}^{2} \right)  \left( {{\it \hat{s}}}^{2}-{{\it \hat{t}}
}^{2} \right) }{{{\it \hat{s}}}^{2}{\it \hat{t}}}}
\nonumber \\ &&  \hspace{-4.0cm}
+4\,\left[ 8\,{{\it \hat{t}}}^{2}{\it \hat{u}}\,{{\it \hat{s}}
}^{2}+{{\it \hat{t}}}^{3}{\it \hat{u}}\,{\it \hat{s}}+2\,{{\it \hat{t}}}^{3}{{\it \hat{s}}}^{2}+{{\it \hat{t}}}^{4}{\it \hat{s}}-{{\it \hat{t}}}^{4}{\it \hat{u}}+8\,{{\it \hat{s}}}^{2}{\it \hat{t}}\,{{\it \hat{u}}}^{2}+4\,{\it \hat{t}}\,
{\it \hat{u}}\,{{\it \hat{s}}}^{3}-{{\it \hat{u}}}^{4}{\it \hat{t}}-3\,{{\it \hat{u}}}^{3}{{\it \hat{t}}}^{2}-3\,{{\it \hat{u}}}^{2}{{\it \hat{t}}}^{3} \right.
\nonumber \\ && \hspace{-4.0cm}
\left. +2\,{{\it \hat{s}}}^{2}{{\it \hat{u}}}^{3}+{\it \hat{s}}\,{{\it \hat{u}}}^{3}{
\it \hat{t}}+{{\it \hat{s}}}^{3}{{\it \hat{u}}}^{2}+{{\it \hat{s}}}^{3}{{\it \hat{t}}}^{2}+{\it \hat{s}}\,{{\it \hat{u}}}^{4}+2\,{\it \hat{s}}\,{{\it \hat{u}}}^{2}{{\it \hat{t}}}^{2} \right] {\it m_f}\,{\rm Bub1} \left( {
\it M_H^2},{{\it m_f}}^{2} \right)
\nonumber \\ && \hspace{-4.0cm} 
\times \left\{{\it \hat{s}}\,{\it \hat{u}}\, \left( {\it \hat{t}}+{\it \hat{s}} \right)  \left( {\it \hat{u}}+{\it \hat{t}} \right) ^{2} \left( {\it \hat{u}}+{\it \hat{s}} \right) \right\}^{-1}
-2\,{\frac {{\it m_f}\,{\it \hat{s}}\,{\rm TriF2} \left( {\it \hat{s}},{{\it m_f}}^{2} \right)  \left( {{\it \hat{t}}}
^{2}+{{\it \hat{u}}}^{2} \right) }{{\it \hat{t}}\,{{\it \hat{u}}}^{2}}} -8\,{\frac { \left( {\it \hat{s}}\,{\it \hat{u}}+{\it \hat{t}}\,{\it \hat{s}}+{\it \hat{t}}\,{\it \hat{u}} \right) {\it m_f}}{ \left( {\it \hat{u}}+{\it \hat{t}} \right) {
\it \hat{s}}\,{\it \hat{u}}}}
\nonumber \\ &&\hspace{-4.0cm}
+2\,
{\it m_f}\, \left( {{\it \hat{t}}}^{3}{\it \hat{u}}\,{\it \hat{s}}+{{\it \hat{t}}}^{4}{\it \hat{s}}+4\,{\it \hat{t}}\,{{\it m_f}}^{2}{{\it \hat{u}}}^{3}-2\,{{\it \hat{u}}}^{3}{{\it \hat{t}}}^{2}-2\,{{\it \hat{u}}}
^{2}{{\it \hat{t}}}^{3}+{\it \hat{s}}\,{{\it \hat{u}}}^{3}{\it \hat{t}}+16\,{{\it \hat{t}}}^{2}{{\it \hat{u}}}^{2}{{\it m_f}}^{2}\right.
\nonumber \\ &&\hspace{-4.0cm}
\left.+4\,{{\it \hat{t}}}^{3}{{\it m_f}}^{2}{\it \hat{u}}+{\it \hat{s}}\,{{\it \hat{u}}}^{4}
 \right) {\rm TriF} \left( {\it \hat{s}},{\it M_H^2},{{\it m_f}}^{2} \right)  \left\{{\it \hat{t}}\, \left( {\it \hat{u}}+{\it \hat{t}} \right) {\it \hat{s}}\,{{\it \hat{u}}}^{2}\right\}^{-1}
 +2\,\left( {{\it \hat{s}}
}^{3}{\it \hat{t}}-{{\it \hat{s}}}^{3}{\it \hat{u}}+{\it \hat{t}}\,{\it \hat{u}}\,{{\it \hat{s}}}^{2}\right.
\nonumber \\ && \hspace{-4.0cm}
\left. -{{\it \hat{u}}}^{2}{{\it \hat{s}}}^{2}+4\,{{\it m_f}}^{2}{\it \hat{u}}\,{{\it \hat{s}}}^{2}-{\it \hat{t}}\,{{\it \hat{u}}}^{2}{
\it \hat{s}}-4\,{{\it m_f}}^{2}{{\it \hat{u}}}^{2}{\it \hat{s}}-{\it \hat{t}}\,{{\it \hat{u}}}^{3} \right) {\it m_f}\,{\rm TriF} \left( {\it \hat{t}},{\it M_H^2},{{\it m_f}}^{2} \right) {{{\it \hat{u}}}^{-2
}{{\it \hat{s}}}^{-2}}
\nonumber \\ &&\hspace{-4.2cm}
+2\,{\frac { \left( -{{\it \hat{s}}}^{3}{\it \hat{t}}+{{\it \hat{s}}}^{3}{\it \hat{u}}+{\it \hat{t}}\,{\it \hat{u}}\,{{\it \hat{s}}}^{2}+4\,{{\it m_f}}^{2}{{\it \hat{s}}}^{2}{\it \hat{t}}-{{\it \hat{t}}}^
{2}{{\it \hat{s}}}^{2}-{{\it \hat{t}}}^{2}{\it \hat{u}}\,{\it \hat{s}}-4\,{{\it m_f}}^{2}{{\it \hat{t}}}^{2}{\it \hat{s}}-{{\it \hat{t}}}^{3}{\it \hat{u}} \right) {\it m_f}\, {\rm TriF} \left( {\it \hat{u}},{\it M_H^2},{{\it m_f}}^{2} \right)}{{\it \hat{t}}\,{\it \hat{u}}\,{{\it \hat{s}}}^{2}}}
;
\end{eqnarray}
\begin{eqnarray}
{\cal G}_4(\hat{t},\hat{u},\hat{s},M_H,m_f) &=& -2\,{\frac {{\it m_f}\,{\it \hat{s}}\, \left( 12\,{{\it m_f}}^{2}{\it \hat{u}}-{{\it \hat{u}}}^{2}+4\,{\it \hat{t}}\,{\it \hat{s}} \right) {\rm BoxT} \left( {\it \hat{s}},{\it \hat{t}},{\it M_H^2},{{\it m_f}}^{2} \right) }{{{\it \hat{u}}}^{3}}}
\nonumber \\ && \hspace{-4.0cm}
+2\,{\frac {{\it m_f}\,{\it \hat{s}}\, \left( 4\,{{\it m_f}}^{2}-{\it \hat{u}} \right) {\rm BoxT} \left( {\it \hat{s}},{\it \hat{u}},{\it M_H^2},{{\it m_f}}^{
2} \right) }{{\it \hat{t}}\,{\it \hat{u}}}}+2\,{\frac {{\it m_f}\, \left( 4\,{{\it m_f}}^{2}-{\it \hat{u}} \right) {\rm BoxT} \left( {\it \hat{t}},{\it \hat{u}},{\it M_H^2},{{\it m_f}}^{2}
 \right) }{{\it \hat{u}}}}
 \nonumber \\ && \hspace{-4.0cm}
 -16\,{\frac { \left( {\it \hat{t}}+2\,{\it \hat{u}} \right) {\it m_f}\,{\it \hat{s}}\,{\rm Bub1} \left( {\it \hat{s}},{{\it m_f}}^{2} \right) }{ \left( {\it \hat{u}}+{\it \hat{t}}
 \right) ^{2}{{\it \hat{u}}}^{2}}}-16\,{\frac {{\it \hat{s}}\, \left( 2\,{\it \hat{u}}+{\it \hat{s}} \right) {\it m_f}\,{\rm Bub1} \left( {\it \hat{t}},{{\it m_f}}^{2} \right) }{ \left( {\it \hat{u}}+
{\it \hat{s}} \right) ^{2}{{\it \hat{u}}}^{2}}}-16\,{\frac {{\it m_f}\, \left( {\it \hat{t}}\,{\it \hat{s}}-{{\it \hat{u}}}^{2} \right) }{{\it \hat{u}}\,{\it \hat{t}}\, \left( {\it \hat{u}}+{\it \hat{s}} \right)  \left( {\it \hat{u}}+{\it \hat{t}} \right) }}
 \nonumber \\ && \hspace{-4.0cm}
+16\,{\frac {{\it m_f}\,{\it \hat{s}}\, \left( 4\,{\it \hat{t}}\,{\it \hat{u}}\,{\it \hat{s}}+2\,{{\it \hat{t}}}^{2}{\it \hat{u}}+{{\it \hat{s}}}^{2}{\it \hat{t}}+{\it \hat{s}}\,{
{\it \hat{t}}}^{2}+2\,{{\it \hat{s}}}^{2}{\it \hat{u}}+5\,{\it \hat{s}}\,{{\it \hat{u}}}^{2}+5\,{\it \hat{t}}\,{{\it \hat{u}}}^{2}+4\,{{\it \hat{u}}}^{3} \right) {\rm Bub1} \left( {\it M_H^2},{{\it m_f}}^{2}
 \right) }{ \left( {\it \hat{u}}+{\it \hat{t}} \right) ^{2}{{\it \hat{u}}}^{2} \left( {\it \hat{u}}+{\it \hat{s}} \right) ^{2}}}
  \nonumber \\ && \hspace{-4.0cm}
 +4\,{\frac {{\it m_f}\, \left( 4\,{{\it \hat{t}}}^{2}{\it \hat{u}}\,{\it \hat{s}}+2
\,{{\it \hat{t}}}^{3}{\it \hat{s}}+2\,{\it \hat{t}}\,{{\it \hat{u}}}^{2}{\it \hat{s}}+{{\it \hat{u}}}^{4}+4\,{{\it m_f}}^{2}{{\it \hat{t}}}^{2}{\it \hat{u}}-{{\it \hat{t}}}^{2}{{\it \hat{u}}}^{2}-4\,{{\it m_f}}^{2}{{\it 
\hat{u}}}^{3}+8\,{{\it m_f}}^{2}{{\it \hat{u}}}^{2}{\it \hat{t}} \right)  }{{\it \hat{t}}\, \left( {\it \hat{u}}+{\it \hat{t}} \right) {{
\it \hat{u}}}^{3}}}
 \nonumber \\ && \hspace{-4.0cm}
\times{\rm TriF} \left( {\it \hat{s}},{\it M_H^2},{{\it m_f}}^{2} \right)
+4\,{\it m_f}\, \left( 2\,{{\it \hat{s}}}^{3}{\it \hat{t}}+4\,{\it \hat{t}}\,{\it \hat{u}}\,{{\it \hat{s}}}^{2}+2\,{\it \hat{t}}\,{{\it \hat{u}}}^{2}{\it \hat{s}}-{{\it \hat{u}}}^{2}{{\it \hat{s}}}^{
2}+{{\it \hat{u}}}^{4}+4\,{{\it m_f}}^{2}{\it \hat{u}}\,{{\it \hat{s}}}^{2}-4\,{{\it m_f}}^{2}{{\it \hat{u}}}^{3}\right. 
\nonumber \\ && \hspace{-4.0cm}
\left.+8\,{{\it m_f}}^{2}{{\it \hat{u}}}^{2}{\it \hat{s}} \right) {\rm TriF} \left( {\it \hat{t}}
,{\it M_H^2},{{\it m_f}}^{2} \right) \left\{\left( {\it \hat{u}}+{\it \hat{s}} \right) {\it \hat{t}}\,{{\it \hat{u}}}^{3}\right\}^{-1}-4\,{\frac { \left( {\it \hat{t}}+{\it \hat{s}} \right)  \left( 4\,{{\it m_f}}^{2}-
{\it \hat{u}} \right) {\it m_f}\, }{{\it \hat{t}}\,{{\it \hat{u}}}^{2}}}
\nonumber \\ && \hspace{-4.0cm}
\times {\rm TriF} \left( {\it \hat{u}},{\it M_H^2},{{\it m_f}}^{2} \right)
-8\,{\frac {{\it m_f}\,{{\it \hat{s}}}^{2}{\rm TriF2}
 \left( {\it \hat{s}},{{\it m_f}}^{2} \right) }{{{\it \hat{u}}}^{3}}}-8\,{\frac {{\it m_f}\,{\it \hat{t}}\,{\it \hat{s}}\,{\rm TriF2} \left( {\it \hat{t}},{{\it m_f}}^{2} \right) }{{{\it \hat{u}}}^{3
}}}.
\end{eqnarray}
Two new integrals appear in these form factors:
\begin{eqnarray}
{\rm TriF2}(p_1^2,m^2) &=& \int \frac{d^d k}{i \pi^{d/2}} \frac{1}{k^2-m^2} \frac{1}{(k+p_1)^2-m^2} \frac{1}{(k+p_1+p_2)^2-m^2} \nonumber \\ 
 && {\rm with} \,\, (p_1+p_2)^2=0,\,\,\,\, p_2^2=0, \nonumber \\ 
{\rm BoxT}(p_1^2,p_2^2,p_3^2,m^2) &=&  \int \frac{d^d k}{i \pi^{d/2}} \frac{1}{k^2-m^2} \frac{1}{(k+p_2)^2-m^2} \frac{1}{(k+p_2-p_3)^2-m^2} \nonumber \\ 
&\times& \frac{1}{(k+p_1+p_2-p_3)^2-m^2} \,\,\,\, {\rm with} \,\, (p_1-p_3)^2=0,\nonumber \\ && (p_2-p_3)^2=0,\,\,\,\,(p_1+p_2-p_3)^2=0.
\end{eqnarray}
It is again useful to study the approximate expressions for these form factors as $m_f$ becomes large.  They take the form
\begin{eqnarray}
{\cal G}_1(\hat{t},\hat{u},\hat{s},M_H,m_f) &\stackrel{m_f \longrightarrow \infty}{\to}& -\frac{4(\hat{u}+\hat{s})(\hat{t}+\hat{s})}{3\,m_f\, \hat{s}\,\hat{u}},\nonumber \\
{\cal G}_4(\hat{t},\hat{u},\hat{s},M_H,m_f) &\stackrel{m_f \longrightarrow \infty}{\to}& \frac{8}{3\,m_f\,\hat{t}}.
\end{eqnarray}

\end{document}